**Predictive Mixing for Density Functional Theory (and other Fixed-Point Problems)**

L. D. Marks

Department of Materials Science and Engineering

Evanston, IL 60201

Northwestern University

**Abstract**

Density functional theory calculations use a significant fraction of current supercomputing time. The resources required scale with the problem size, internal workings of the code and the number of iterations to convergence, the latter being controlled by what is called mixing. This note describes a new approach to handling trust-regions within these and other fixed-point problems. Rather than adjusting the trust-region based upon improvement, the prior steps are used to estimate what the parameters and trust-regions should be, effectively estimating the optimal Polyak step from the prior history. Detailed results are shown for eight structures using both the "Good" and "Bad" Multisecant versions as well as Anderson and a hybrid approach, all with the same predictive method. Additional comparisons are made for thirty-six cases with fixed algorithm Greed The predictive method works well independent of which method is used for the candidate step, and is capable of adapting to different problem types particularly when coupled with the hybrid approach. It would be premature to claim that it is the best possible approach, but the results suggest that it may be.

# 1. Introduction

A classic problem in many different areas of science is solving a set of non-linear equations, what is called solving a fixed-point problem. One area where this very important is in density functional theory (DFT), where the solution to the Kohn-Sham equations[1,2] are a fixed-point density under the action of a non-linear quantum mechanical operation. The total time (and computational effort) scales as the multiple of the time for each individual step, for instance solving the eigenvalue problem, and the total number of steps. The latter is determined by how efficiently the fixed-point problem is solved.

While solving fixed-point problems (root finding and/or non-linear equations) is described in many undergraduate courses on numerical methods, they are not as simple as they might seem. This is particularly the case when gradients are either not available or are prohibitively expensive to calculate, as in DFT problems. The most common approach for these is to use quasi-Newton methods or a least squares approach to approximate the Jacobian – the two are mathematically equivalent to second order. This is similar to methods used for optimization where the mathematical background and methods are much better developed. Unfortunately, fixed-point problems do not necessarily possess the local convexity of optimization problems, and convergence is not so readily guaranteed. Indeed, it is really somewhat amazing that DFT problems with thousands of variables often converge in only a few tens of self-consistent iterations. If they did not converge that fast, modern ab-initio methods would be considerably less useful.

It is established that one can obtain convergence of the Kohn-Sham equations with appropriately chosen parameters for the self-consistent iterations and exact functionals, e.g.[3-9], although whether this will occur in finite computational time is not guaranteed. Unfortunately, convergence with general parameters, imperfect functionals and starting points far from the solution in the hands of inexperienced users of black-box codes is not guaranteed. While convergence can often be obtained by adjusting internal parameters until they work, this is highly undesirable; on many DFT listservs the most common question is a request for assistance on how to converge a calculation. A target has to be a method that will converge fast without needing user intervention.

Why do the calculations sometimes not converge? In the DFT literature how they behave has often been analyzed in terms of physical variations of the electron density. A classic example of this is for large, metallic calculations, for instance for a surface. In such cases the density can oscillate across the material, for instance between the outer surface and the inside. This phenomenon is called "sloshing" in the DFT literature, and the oscillations can increase in magnitude leading to divergent behavior. This would be described differently mathematically, namely as a consequence of the algorithm taking steps that are larger than appropriate, with positive feedback leading to amplified oscillations. Both interpretations are correct, but the physical interpretation in general does not provide enough guidance on how to avoid this issue whereas the mathematical one does.

More critically, in some cases the physical interpretations can be misleading and have led to "scientific myths" (e.g.[10] for an informative analysis). These are ideas or concepts that seem to be reasonable, are widely used, but in fact are fundamentally flawed. One of the most common of these is the idea that instabilities such as the sloshing mentioned above can be avoided by reducing what is called the "mixing factor", more rigorously the algorithm Greed as will be discussed later. While this might work, it is not correct mathematics so can fail catastrophically. At least in the opinion of the author solving such issues requires treating equally the mathematics and physics (or chemistry) of the problem.

It is established in optimization and general methods for non-linear equations (e.g.[11-15]) that convergence can only be guaranteed if either a line search or a trust region is used; for more details on trust region methods see Yuan[16]. In most cases a trust region approach is faster than a line search for DFT as it requires a smaller number of computationally (very) expensive calculations. In optimization, how large the trust radius should be is well defined because in general gradients are available; if the function reduces significantly it is good so they can be expanded, the opposite if the function does not reduce as expected. In energy-based DFT methods where true gradients are available, trust regions are common, e.g. [17-21]. However, energy minimization is fundamentally different from a fixed-point method without gradients.

While how to handle trust regions is well established in optimization, it is not for fixed-point problems where there can be problems with scaling and other uncertainties. What is a reasonable reduction to use to expand the trust radius, particularly when the problem can change discontinuously for something as simple as an isolated oxygen molecule (e.g.[22, 23]), or when during convergence the system transitions from metallic to insulation? How should one scale components such as the kinetic energy density in metaGGA methods, or other terms? In addition, gradients are not available so most standard trust region methods such as Dogleg[24] are not appropriate.

The intention of this note is to suggest a different approach to handle trust regions and control of parameters in fixed-point methods, focusing on DFT although the method should be general. Instead of contracting/enlarging these based upon some potentially biased metric, the approach calculates what should have been used for the most recent step and then uses this as a prediction of what to use in the current step – effectively calculating the optimum Polyak step size[15]. The advantage of this approach is that it removes the need for the user (or programmer) to decide what is good and what is not, instead this is determined automatically. This predictive approach is not specific to one particular method of choosing the candidate step for the next iteration, and is completely general. There is no need to have a set of parameters appropriate for a metal, a different one for a surface and a third set for an insulator.

The structure of this paper is as follows. In the next section I will recap the basic structure of linear mixing methods, paying attention to points that have often been either glossed over or ignored in the literature, but are important for production code. Some aspects of this are specialized to all-electron codes such as Wien2k[25, 26], where they have been implemented[22, 23, 27] in the production version. However, they are general and much of this section is presented so other codes can use appropriate scaling and trust region methods in their mixing algorithms, avoiding pitfalls. Some of these are were described in earlier work[22, 23, 27]; some have been mentioned in the "Readme" notes with different Wien2k releases, but never described in detail; some exist in the literature outside DFT; significant parts of the analysis and algorithm are new.

I will then explain the predictive algorithm. This is followed first by examples comparing different algorithms commonly used in DFT calculations showing how the step sizes and trust regions change for different underlying algorithm. This is followed by a more detailed analysis comparing the predictive approach to one without most of the controls for thirty-six structures across a wide range of different systems.

While it would be extreme to claim that the predictive approach described herein will solve all problems, it appears to come close. Tests to date both by the author and by others indicates that it does not as yet fail provided that the underlying physical problem is decently posed.

## 2. Linear Mixing Methods

The purpose of this section is to recap the basics of mixing, and provide updates to some aspects of the algorithms developed previously which have not been formally published although they have been present in the production code for Wien2k[25, 26] for some years. At the same time the similarity and differences between the original Broyden[28] approach, the commonly used Anderson acceleration[29-35] which is sometimes called the Pulay method[36-40] and sometimes Direct Inversion in the Iterative Subspace (DIIS), as well as scaled multisecant[22, 23] and a hybrid method[27] will be described. The Anderson, Pulay and DIIS methods are nominally the same, although they could be different in terms of how they are implemented numerically where details are often not published.

It is important to recognize that there are three essential components to a mixing algorithm, all of which need to be defined for reproducibility:

1. What the variable are, including scaling and preconditioning, as well as metrics.
2. Determination of the candidate step, including both the predicted and unpredicted components[22, 23] as well as regularization.
3. Control of the step used, for instance a line search or trust region method.

It is common to find mixing methods defined just by stating the method of determining the candidate step, whether it is DIIS, Pulay or other, and incomplete details. This is equivalent to stating that DFT calculations were performed without defining the functional used.

### 2.1 Basic Formulation

The formalism for mixing is well established, and will be described here for consistency based upon prior work of the author[22, 23, 27] in order to form the basis for some of the other analysis in this and other sections; a recent analysis with different opinions can be found in references[41, 42]. Consider some density $\rho(r,R)$ as a function of position $r$, and a vector of $N_a$ atom positions $R = (R_1, R_2, ... R_m)$. The solutions of the equations of Kohn-Sham density-functional theory[1, 2] can be written as:

$$(H_0 + V)\phi_i = \epsilon_i \phi_i \tag{1}$$

$$F(\rho(r,R)) = \sum_i (1 + exp((\epsilon_i - \mu)/kT))^{-1}|\phi_i|^2 \tag{2}$$

with eigenvectors (orbitals) $\phi_i$ and eigenvalues (energies) $\epsilon_i$, where $H_0$ is the noninteracting single-particle Hamiltonian, $V$ the effective local potential, $\mu$ the chemical potential, $k$ Boltzmann's constant and $T$ the temperature. In its simplest form we seek the fixed-point solution for a given density and atomic positions, i.e.

$$F(\rho(r,R)) - \rho(r,R) = D(\rho(r,R)) = 0 \tag{3}$$

where $D(\rho(r,R))$ is the density residual which is analogous to the negative of the gradient in optimization. This form is specific to pure DFT calculations where the only active variables are the density; the forms for orbital-based density methods are slightly different and would involve a density matrix, an orbital potential or include the wavefunctions as part of the active variables. These, as well as other terms such as constraints on atomic positions or linearization energies can all be included without problem. In addition to the density condition, for a complete solution we seek the minimum of the total energy of the system, i.e. for a total energy $E(\rho(r,R))$ (including electronic entropy[2]) the derivative as a function of the atomic positions:

$$\partial E(\rho(r,R))/\partial R_i = g_i(\rho(r,R)) = 0 \qquad (4)$$

Many algorithms solve equation (3) for some fixed atomic positions, then change them using a minimization algorithm to move towards the solution of equation (4), reconverge the density and iterate. Rather than solving these serially, they can be merged, i.e. we seek the fixed-point solution of

$$D(\rho(r,R)), -g(\rho(r,R)) = Res(\rho(r,R)) = 0 \qquad (5)$$

with $Res(\rho(r,R))$ a generalized residual vector. This is an approach initially suggested by Bendt and Zunger[43], and has been implemented[27] within the Wien2k code. For a DFT code where the basis set is not atom position dependent the gradients are the negative of the Hellmann-Feynman forces; otherwise additional corrections for the basis set are needed, commonly called Pulay corrections[44-46]. These are calculated with the Kohn-Sham density of $F(\rho(r,R)$ whereas the Hellmann-Feynman calculations use $\rho(r,R)$; hence the gradients $g(\rho(r,R))$ in (4) and (5) are not the Born-Oppenheimer surface gradients, they are instead vectors that share a common fixed-point. (The Born-Oppenheimer surface is defined as the energy as a function of $R$ for $D(\rho(r,R)) = 0$.)

Both the approach and notation used to solve equations (3) or (5) varies. Some algorithms deal with them as a least squares problem, others use a Taylor series – herein I will use the latter. As has been demonstrated previously by several authors the approaches are nominally identical (e.g. [11, 22, 23, 47-50]), but differ in important implementation details. Dropping the $(r,R)$ notation for brevity, and using $\rho_n$ and $R_n$ to describe the density and positions respectively for iteration 'n' as well as $G_n$, this suggests a standard Newton algorithm using a vector/matrix representation:

$$(\rho_{n+1}, R_{n+1}) = (\rho_n, R_n) - H_n Res_n + O(Res_n^2) \qquad (6)$$

$$Res_{n+1} = Res_n - B_n \Delta(\rho_n, R_n) + O(\Delta(\rho_n, R_n)^2) \qquad (7)$$

where $H_n$ is the inverse of the Jacobian $B_n$ for the change in density/pseudo-gradients with density/atomic positions. (The use of "$H$" here is conventional, unfortunately confusable with a Hamiltonian.) When terms of order $Res_n^2$ or $D_n^2$ in the Taylor series are neglected this is a linear model; the neglect of these terms is important as will be shown later. The computational cost of calculating the complete $H_n$ or $B_n$ is prohibitive, so instead they are approximated using a quasi-Newton method. Introducing the new variables:

$$y_{j,n} = Res_n - Res_j \qquad (8)$$

$$s_{j,n} = (\rho_j, R_j) - (\rho_n, R_n) \tag{9}$$

and the matrices $S_n = [s_{n-k,n}, s_{n-k+1,n}, \ldots s_{n-1,n}]$ and $Y_n = [y_{n-k,n}, y_{n-k+1,n}, \ldots y_{n-1,n}]$, we require that $H_n$ and $B_n$ satisfy the multisecant equations:

$$H_n Y_n = S_n \text{ or } B_n S_n = Y_n \tag{10}$$

for which there are general rank-one solutions

$$H_n = \sigma_n[I - Y_n Inv(Y_n^T W)W^T] + \beta_n S_n Inv(Y_n^T W)W^T \tag{11}$$

$$B_n = \eta_n[I - S_n Inv(S_n^T V)V^T] + \gamma_n Y_n Inv(S_n^T V)V^T \tag{12}$$

with $W$ and $V$ any vector of size $N_b + 3N_a$ (the size of the basis set and the number of atomic positions variables), $\sigma_n$ and $\eta_n$ are the *algorithm Greed*, $\beta_n$ and $\gamma_n$ will be referred to as *Damping terms* included to account for the neglect of the higher order terms in equations (6) and (7), and $Inv$ stands for inverse which will be discussed in §2.3. As discussed in earlier work[22, 23, 27], the Greed determines how large a step the algorithm will take in the unpredicted direction and plays a critical role. A definition of "Greedy algorithms" is appropriate[51]:

*"A Greedy algorithm always makes the choice that looks best at the moment. That is, it makes a locally optimal choice in the hope that this choice will lead to a globally optimal solution."*

Too much Greed can lead to instability depending upon the magnitude of the higher-order terms in equations (6) and (7); too little and the algorithm may not converge or only very slowly. (The latter case where steps are too small has analogies to the standard Wolfe and Armijo-Goldstein conditions[13].)

In addition to the Greed and non-linearity control via the Damping, one other term is useful to define. If $\|Y^T Y\|_F \ll \|S^T S\|_F$, where $\|A\|_F$ is the Froebius norm of the matrix $A$, the problem can be defined as "soft" in the spirit of soft phonon modes or similar in a phase transition. In contrast, if the inequality is in the other direction the problem is "hard". Hard problems may also be "stiff" in the sense used for differential equations, although this does not have to be the case.

The next step is then given by

$$(\rho_{n+1}, R_{n+1}) = (\rho_n, R_n) + H_n Res_n = (\rho_n, R_n) + \sigma_n U_n + \beta_n P_n \tag{13}$$

$$U_n = [I - Y_n Inv(Y_n^T W)W^T]Res_n; \; P_n = S_n Inv(Y_n^T W)W^T Res_n \tag{14}$$

Here $U_n$ is the component of the current residue about which no information is available, the unpredicted part; $P_n$ satisfies the secant conditions of equation (10) and is the predicted part based upon previous steps. While in principle both $S_n$ and $Y_n$ can include all previous densities, in practice only a small number, between 6-16 is needed. (Some DFT codes report a need to use much larger numbers which may be due to bad scaling or regularization; the number should not be significantly larger than the number of important eigenvectors of the Jacobian for the current residue.)

These equations deliberately do not specify the form of U or W; I will return to these after some other key components are described.

## 2.2 Units

The formulation described in the previous section is deceptively simple. While it can be (and often has been) implemented as written with whatever parameters are most convenient, this is not optimal. Significant improvements can be achieved by ensuring that the variables are consistent with both the underlying mathematics and physics. This point was made in 1988 by Blügel[52], but was missed at least for earlier versions of the Wien2k code; it is certainly omitted in some public domain codes.

The first issue that needs to be consider is units. Equations (6)-(14) use standard vector/matrix representation, which implicitly leads to L2 metrics for the matrix elements of (for instance) $Y_n^T W$. These metrics have to be physically consistent. In particular, the contributions from different parts of the basis set has to be invariant to changes such as doubling of the unit cell, as well as reducing the symmetry of Fourier or local contributions, and also non-density parameters such as density matrix or orbital potentials. They also have to be invariant to, for instance, increasing the volume of space without electrons in a surface calculation or for an isolated atom.

To illustrate this, consider an isolated atom with electrons inside the muffin tin and also plane waves in a cubic unit cell of volume V, where the volume is large, and ignore the plane wave pseudo-charge inside the muffin tins (see Appendix 1). The change during mixing in terms of electrons per unit volume is not independent of the volume, so is inappropriate. Units of total electron to some power for the vector products in $Y_n^T W$ is independent of the volume, so products which are proportional to this are viable. Noting that the number of plane waves to some maximum wavevector scales with the volume, whereas the densities scale inversely with the volume, the volume integral of the density squared when the number of planes wave is included scales as the total number of electrons squared. If one now changes the size of the muffin tins to transfer some density from those to the plane waves, it follows that the same units are required for the products within the muffin tins.

Hence plausible units of the vector products in $Y_n^T W$ are the total charge squared or the volume integral of density squared summed over all variables. Tests indicate that the latter is better, consistent with the fact that we are dealing with "density" functional theory. Being specific, for a vector of variables $A(r,R)$ the L2 squared should be the appropriate integral over volume. i.e.

$$(L2)^2 = \|A\|^2 \equiv \int w(r) A(r,R)^* A(r,R) \, dV \qquad (15)$$

with $w(r)$ the appropriate scaling of input variables which includes the multiplicity of atoms and/or plane waves and other terms. Note that this is *not necessarily* the same as a density cross-product since components of the basis set do not have to be orthogonal. The fixed-point problem is solved for the variables used which includes the pseudo-charge (Appendix 1) and other terms. The units for $(L2)^2$ is then electrons squared per unit volume. Scaling is done by preconditioning the vectors and matrices by multiplying them by $\sqrt{w(r)}$ to later exploit BLAS/LAPACK calls, removing this scaling at the end.

Less clear is how to scale orbital terms such as density matrices or atomic positions (except for multiplicities which enter by a square root similar to equation (15)). In the original implementations[22, 23, 27] an adjustable scale was used, but it now seems best to use fixed scales. Atomic positions are used in atomic units (au), pseudo-forces in Rydberg/au, and energy units are Rydberg, all divided by $\sqrt{4\pi}$ which is an adequate scaling.

## 2.3 Regularization

The equations in §2.1 implicitly involve least-squares fits, and implicitly or explicitly include a matrix inverse, which can lead to ill-conditioning, see for instance[14, 53, 54]. Consequently, all numerical mixing approaches involve some form of regularization, which may be machine precision truncation or some default in library codes. The majority of literature analyses do not state the regularization used.

Experience with the Wien2k code indicates that a conventional Tikhonov regularization[55] of the singular values of $W^T Y$ by $\lambda = \xi_{max} * 2 \times 10^{-4}$ is approximately correct, where $\xi_{max}$ is the largest singular value. Significantly smaller values such as $\xi_{max} * 1 \times 10^{-8}$ can lead to instabilities, larger ones such as $\xi_{max} * 1 \times 10^{-2}$ decrease the speed. Although methods to estimate the regularization exist, e.g.[56, 57], they do not seem to work in tests. Being specific, with the standard single-value decomposition of a matrix $A$ as

$$A = U \Sigma V^T \text{ with } \Sigma_{ii} = \xi_i \tag{16}$$

the *Inv* operator in equations (11) is interpreted as

$$Inv(A) = V \Theta U^T \text{ with } \Theta_{ii} = \xi_i / (\xi_i^2 + \lambda^2) \tag{17}$$

In the limit of no regularization, with $W=V$ equation (11) and (12) are true inverses; with regularization, they are not and the term "dual" is more appropriate. An extended analysis of other types of regularization can be found in reference[32]

One useful metric is the effective rank, defined for M memories as

$$Rank = (1/M) \sum_i \xi_i^2 / (\xi_i^2 + \lambda^2) \tag{18}$$

Empirically, values of 0.7-0.9 indicate that the problem is fairly well-posed; both larger and smaller indicate potential issues.

## 2.4 Non-Linearities and Trust Region

A standard phenomenon with DFT calculations, as mentioned earlier, is what is often called "charge sloshing", where density appears to oscillate between different regions. It is important to understand the underlying mathematics.

All methods (except the Tensor approach[58-62]) ignore the higher-order terms in equations (6) and (7). Provided that these are small, solutions to equation (6) can be expected to converge superlinearly (see[11, 13, 14, 22, 23, 27] and references therein). However, it is not appropriate to ignore the higher order terms, particularly far from the solution. Because of these non-linear terms there are (at least) two mathematical sources of sloshing. The first is the magnitude of the unpredicted step in equation (11). Depending upon the problem (the inverse Jacobian along the unpredicted direction) this should be either small or large. The second source for is the neglect of higher-order terms in the predicted step. The Damping terms $\beta_n$ and $\gamma_n$ account for this, and are both problem dependent and change as the variables converge to the fixed point.

The standard approach for quasi-Newton methods to handle such issues is to limit the step size, both that along the predicted and unpredicted directions, using what is called a Trust Region (e.g.[11, 13, 15, 16]); this is comparable to how stiff differential equations have to be handled. A trust region radius is controlled inside the code, and candidate steps are limited so that they are not larger than this radius. If the calculation is making good progress the trust radius is expanded; if it is not the trust radius is decreased. As illustrated in Figure 1, depending upon the sign and magnitude of the ignored terms, controlling non-linearities will require a small trust radius or a large one – it is highly problem dependent. For the curve "1" in the Figure the non-linearities are not critical, but for curve "2" and "3" which have positive and negative second-order terms respectively, not only is their over-shooting of the "best" value, but the Simplex gradient[63] that would be used in future steps are poor. (The set of prior directions in the Simplex gradient[63] has many of the properties of regular gradients[64], but also differences as they are a limited set of directional secant slopes with finite step sizes.)

Note also from Figure 1 that the optimal step along the unknown direction ($\sigma_n$ of equation (13)) has to be large enough to come close to the minimum residual along this direction, but also not too large. The optimal step is often called the Polyak step[15], a convention that will be used herein. With too small a value of the Greed the algorithm can "starve to death"; this shows up in calculations as will be seen in §5.6; too small a value also often leads to premature convergence if the atomic positions are within the fixed-point method. The need for a step which is not too small has similarities to the standard Armijo-Goldstein inequality for sufficient decrease, and similar step controls used in optimization[12, 13].

For the predicted component the trust-region approach used in the Wien2k code is a standard method, solving the subproblem that minimizes the Lagrangian $L$ for a step which is a linear combination ($V$) of the prior steps, of size $C$

$$L = \|B_n V S_n - \beta_n B_n P_n\|^2 - \psi(\|V S_n\|^2 - C^2) \tag{19}$$

And then using for the next step

$$(\rho_{n+1}, R_{n+1}) = (\rho_n, R_n) + \sigma_n U_n \|V S_n\|/\beta \|P_n\| + V S_n \tag{20}$$

reducing the size by changing $\psi$ until the trust region conditions are obeyed. This is equivalent to a Levenberg-Marquandt[65, 66] algorithm for the predicted step; the unpredicted component is scaled down by the reduction of the predicted component. Note that equation (20) solves for the best residue decrease, as against the step closest to a full step.

For the unpredicted component, earlier versions in Wien2k used an implicit trust approach where the size of the Greed $\sigma_n$ was increased/decreased depending upon whether the algorithm was making progress. The total step and other metrics were also controlled by a trust radius which increased when the algorithm was making progress, decreased when it was not. This worked, although there is a better approach as will be described later. For completeness, a slightly different form of the implicit trust region for the Greed has been implemented in the Castep code[67].

## 2.5 Backtracking

No matter how well protected a mixing (or optimization) trust region algorithm is, bad steps will always occur. While the next step may recover, a single bad step contains misleading information and will contaminate the Simplex gradients. The standard approach is backtracking, that is to return to the prior point, then re-evaluate what should be done. The current algorithm uses one of two approaches depending upon the residue increase:

1. If the increase is larger than a number $N_1$ (default 2.0), performs a quadratic fit along the prior step and then use this, discarding the most recent step (so it does not contaminate the Simplex gradients of future iterations).
2. If the increase is smaller than $N_1$, but larger than $N_2$ (default 1.5), keep the current information but go back to the previous position and residue and recalculate after repeating the prediction for the trust radius and other terms (see §3). Note that this temporarily increases the number of memories used by one.

Extensive tests indicate that the quadratic fit and ignoring the most recent value is better than keeping it. The two defaults are approximately correct, and are not dependent upon the type of system being analyzed as they are limiting contamination.

## 2.6 Choice of Algorithm and History Scaling

The units, regularization, trust region control and backtracking in sections 2.2-2.5 do not depend upon the specifics of how the matrices $W$ and $V$ in equations (11) and (12) are defined. The main algorithms are summarized in Table 1, although results are not presented for all of them herein and this table may not be complete.

The original algorithms suggested by Broyden[28] used a sequence of rank-one updates, rather than a matrix about the current point, satisfying the secant equations for a series of steps. Since the most recent information overwrites earlier information this is a greedy algorithm. While this approach was used in earlier DFT codes[52, 68, 69], it is no longer common with multisecant matrix methods dominant, i.e. the matrix secant equation (10).

More common in the DFT literature is to center the matrices on the current point. Different forms for $W$ and $V$ and slightly different methods of constructing the mathematics lead to variations in the method; all the common methods used for DFT can be described in this fashion as discussed previously[23]. In the mathematics literature the two most common are to take $W=S_n$, a multisecant form of Broyden's first method, often called "good Broyden", whereas taking $W=Y_n$ is a multisecant form of his second method, sometimes called "bad Broyden". In the original paper by Broyden[28] where non-multisecant methods were introduced his first method worked, his second did not. A number of papers soon afterwards reached the same conclusion, so "bad Broyden" was largely dismissed. However, more recent work in the mathematics literature have questioned whether the "bad" methods fails in all cases.

One common variant is what is called Anderson acceleration[29-35], the Pulay method[36-40] and sometimes Direct Inversion in the Iterative Subspace (DIIS) – they are all nominally the same and implicitly are centered on the current point (not consecutive points). These use $W=Y_n$ without any scaling. The conversion from equation (10) to the multisecants of equations (11) and (12) are implicitly least-squares fits over the prior history. One problem that is rarely mentioned is that least-squares are biased towards the largest residue (or step), which is different from the sequential approach where relative scaling is not an issue.

To avoid this bias, the approach introduced previously[23] is to rescale by the diagonal values of $sqrt(Y_n^T Y_n)$. This removes the bias and also improves the conditioning of the inverse in the equation. This is the MSEC algorithm that has been used since 2008 in the Wien2k code. The same scaling can be applied to the multisecant version of "good Broyden", which I will refer to as MSGB. As discussed previously[23], MSGB is a much more greedy algorithm.

A different approach introduced in 2013 is the MSR1 algorithm[27] in the Wien2k code[25, 26], which is inspired by the symmetric rank one (SR1) optimization algorithm[13, 70-73]. It uses a linear combination of MSEC and MSGB, i.e.

$$W_n = V_n = Y_n + \alpha_n S_n \tag{21}$$

This algorithm is a member of the fixed-point Broyden family, and will share the convergence properties of similar algorithms (e.g.[53, 54, 74-85]). In the limit $\alpha_n = 0$ it is equivalent to Broyden's second method which most DFT codes use; in the limit $\alpha_n \to \infty$ it becomes Broyden's first method.

To bound the value of $\alpha_n$ two limits are applied (not the same as those used originally):

a) An approach similar to that of Shanno and Phua[86] for initial scaling in quasi-Newton methods, namely an upper bound

$$\alpha_n = max(1.0, \|Y^T Y\|_F / \|Y^T S\|_F) \tag{22}$$

For a problem where $\|Y^T Y\|_F \gg \|Y^T S\|_F$ this includes some of the good-Broyden behavior, but not too much. For a soft problem, it tends towards good-Broyden.

b) The largest value of $\alpha_n$ which is less than or equal to that in equation (22), for which $Y_n^T W_n$ has no negative eigenvalues, with the imaginary component less than the real part based upon extensive tests[87]. To explain this, consider some residual $Res$ multiplying the inverse Jacobian of equation (11). We can decompose this vector into three parts:

$$Res = Ya + Sb + Z \text{ with } Z^T Y = Z^T S = 0 \tag{23}$$

and $a$ and $b$ vectors of size of the memory used, whose precise values do not matter here. Looking at the parts:

1. The contribution of $Z$ is independent of $\alpha_n$ so it can be ignored.
2. The component $Ya$ satisfies the secant (least squares) condition, so there is no restriction on it.

3. For the component $Sb$ the action of $H_n$ would be equivalent to a non-positive-definite matrix if there are negative eigenvalues. It is established (e.g. discussion in[27]) that the Jacobian and its inverse are related to the dielectric band structure[88-94] which has positive definite eigenvalues at its solution. Hence the condition b) enforces a minimization structure for the components beyond the secant condition, similar to what is used for optimization problems (e.g.[13]). Here there are similarities to the SR1 algorithm.

For scaling, each memory is scaled such that $Y_n^T(Y_n + \alpha_n S_n)$ has unitary diagonal values – this is better than the prior scaling such that $Y_n^T Y_n$ has unitary diagonal values. Note that this requires a simultaneous rescaling and iterative solution for $\alpha_n$, and will improve the condition number of the inverse.

One point should be made about the difference between MSGB, MSEC (or DIIS) and MSR1. Consider the vector space spanned by non-zero predicted components from equation (14), i.e. the set of vectors $\Psi$ which satisfy $W^T \Psi \neq 0$. For MSGB this vector space is that of the prior $S_n$; for MSEC and DIIS it is the vector space of the prior $Y_n$; for MSR1 it is the sum of the two spaces. This implies that the predicted fraction of any residue component in MSR1 will in general be larger. Note that from b) above the extra prediction (relative to MSEC/DIIS) is connected to a positive definite Jacobian which is appropriate near the solution.

As expected, both MSEC and DIIS can have difficulty with soft problems. Whenever a least squares problem is solved such as that implicit in the inverse of equations (11), the smallest eigenvalues (or singular values) are either discarded or regularized; the stability of each eigenvector is determined by the ratio of its eigenvalue to the largest eigenvalue. Hence with these methods, smaller eigenvectors in the space of the residual and Y are damped. In contrast, MSGB solves a least squares problem in the other space, so implicitly includes more of the small eigenvalues in the space of the residual – sometimes too much which is why it can be unstable. MSR1 is between the two extremes, and hence does significantly better for problems with soft modes where MSEC and DIIS can fail.

In tests, MSR1 is the fastest and most robust, and can also avoid local stagnation that can occur with other methods; MSEC and DIIS are useful but can have problems being not greedy enough; with the predictive mixing that will be described next MSGB does work and is often competitive, but can be noisy as it is often too greedy.

## 3 Predictive Mixing

Mixing and fixed-point methods without some of the details discussed in the previous sections often work, but can fail or require very small trust radii or Greed and therefore be very slow. The target should be a method that always works, does not require user intervention, and should be fast. Earlier versions of the code[22, 23, 27] used the L2 of the residue as a metric on whether to increase or decrease the trust radii, and also change the Greed and predictive step scalings $\sigma_n$ and $\beta_n$. This is the standard approach (e.g.[11, 13]). However, there are also some fundamental problems/issues with using the L2 of the residue:

1. It is unclear how valid an L2 metric of the residue is since some of the variables may be poorly scaled (and/or ill-conditioned). This will vary with problem and is hard to handle except by undesirable adjustment on a case by case basis, e.g. one set of parameters for metals, another for insulators.
2. There are ambiguities as to the units to use – relative to the current residue or absolute?
3. During the iterations the problem often changes (electronic phase transitions[27]), and the trust radii may then change significantly. While it is possible (common) for the user to delete the prior history and restart (e.g.[11-13, 36, 85, 95, 96]), I view this as undesirable for production code.
4. The parameters have to be fundamentally different for stiff problems, where both the control parameters and the trust region have to be small, and soft problems where they need to be large. If the trust radii are not large for soft problems then false convergence will be obtained in simultaneous minimization of atoms and density. It sometimes occurs that the problem is stiff far from the solution, but soft near it. The program, not the user, has to be able to handle this.
5. Some commonly used components of trust region methods such as Dogleg[24] or Cauchy steps[13] are not reliable as no gradient is available, and the residual may not be a good descent direction. In many cases the angle between the residue and step (e.g. equation (20)) is large, sometimes close to ninety degrees; in such cases the residue almost certainly is not close to the steepest descent direction.

Is there a better method? I will argue here that there is. Consider the general form for the next step as

$$H_n Res_n = \sigma_n U_n + \beta_n P_n \tag{24}$$

We can estimate both $\sigma_n$ and $\beta_n$ by calculating what they should have been for the *prior* step, i.e. we minimize for $\sigma_n$ and $\beta_n$ using the *current* Jacobian, i.e. minimize

$$\|U_{n-1} - \sigma_n B_n U_{n-1}\|^2 \text{ and } \|Y^P_{n-1} - \beta_n B_n P_{n-1}\|^2 \tag{25}$$

where $Y^P_{n-1}$ is the projected component of the prior predicted residue. In some case the solutions for both $\sigma_n$ and $\beta_n$ are negative; only the absolute value is used.

Once values for both $\sigma_n$ and $\beta_n$ have been calculating, the maximum step size that should have been used for the last iteration, $\|\sigma_n U_{n-1} + \beta_n P_{n-1}\|$, is calculated and used to estimate the new trust radius, again with a running average to avoid severe fluctuations. For cases where atomic positions are simultaneously converged a trust region control on the atomic step is used based upon this maximum step, that is the maximum allowed atomic step corresponds to the maximum atomic motion in $\|\sigma_n U_{n-1} + \beta_n P_{n-1}\|$. These are used as the trust region radii described in §2.4. The same approach is used for MSR1, DIIS, MSGB and MSEC; the only difference is the matrices used.

To briefly expand, instead of increasing or decreasing the parameters $\sigma_n$ and $\beta_n$ depending upon whether the residue improves or not, they are changed based upon what they should have been for optimal performance in the last iteration. This is equivalent to estimating the Polyak step[15] involving the different parameters. These values are also used to determine the trust region radii of the total step and also the maximum atomic motion . This exploits the implicit assumption in all quasi-Newton methods that the prior history of steps is an adequate predictor. It should be noted that this has similarities to the approach used in tensor methods[58, 59] where the previous information is used to estimate a higher-order term, the initial scaling approach of Shanno and Phua[86] in optimization as well as what is called the BB two-point step size gradient method[97-101]. However, it is not the same and appears to be a useful and new alternate to conventional trust region approaches.

One addition point deserves mention: should the Greed and Total Size be in absolute units or relative? To sidestep this an average over the relative and absolute values appears reasonable in practice. The values tend to behave as relative, although this is determined by the solutions of equation (25) and is not specified in the algorithm.

## 4. Algorithm Summary

The same predictive algorithm is used for the different values of the matrix $W$ (Table 1) for all of MSR1, DIIS, MSGB and MSEC. (It has also been tested for the sequential methods, although in tests these are inferior so will not be discussed further.) The algorithm can be summarized as:

1) Collect prior histories and create the vectors $Y_n$ and $S_n$ for a user specified maximum number of prior values as described in §2.1, converting units as described in §2.2. The default is 8 prior values for just density values, 10 when atomic positions are simultaneously being optimized.
2) If the step was not good, either backtrack or recalculate as described in §2.5.
3) Rescale as described in §2.6, depending upon which algorithm is being used, and also regularize following §2.3
4) Calculate the Greed and Damping parameters as described in §3 for the specific algorithm, and also the maximum step and atomic movement which are used in 6) below as the trust region radii.
5) Calculate the candidate step using these Greed and Damping values for the specific algorithm.
6) If the step or atomic movement are too large reduce them via a trust region approach as described in §2.4
7) Unpack the variables and calculate using the new density, atomic positions, orbital potentials and other parameters.

For the very first iteration a small Pratt-like step with a Greed typically around 0.035 is used – the initial value is chosen based upon the initial residue as described previously[27]; the exact value is not important so long as it is not too small or too large. The predictions are used after the first Pratt-like step, with no rescaling performed until there are two or more memories.

The only adjustable parameters in the algorithm are:

1) The number of memory steps, typically 8-12.
2) The Greed in the first iteration, which just needs to be small enough.
3) The maximum step in the first iteration, which rarely matters.
4) The regularization, although the default value appears to be adequate.
5) When to turn on backtracking.

The results are weakly sensitive to these; none of them are adjusted in any of the results herein, with the exception of the regularization for DIIS in Figure 2b. The most important parameters are determined by the code during the iterations, specifically the Greed, Damping, trust radius size and maximum atomic movement. There is no need to adjust anything for different materials whether they are molecules, insulators, metals, contain transition metals, lanthanides, involve spin-orbit, van-der Waals terms, moving atoms, Hubbard U as well as on-site or full hybrids (and combinations of these).

## 5. Results

In the next sections results will be presented for MSR1, MSEC, DIIS, MSGB (see Table 1); Crystallographic Information Files (CIF) for all are included in the supplemental material. In all cases the same regularization, predictive algorithms for the total step size, Greed and Damping and initial parameters were used, with only the form of $W_n$ and preconditioning changing.

Results herein can in general be broken into five classes:

a. Simple problems such as bulk unit cells, which typically converge in 10-12 iterations starting from a superimposition of isolated atom densities with all the algorithms. Here all that is required is reasonable trust controls to avoid too much Greed.
b. Simple problems where the atomic positions and densities are simultaneously optimized, such as for bulk $Mg(OH)_2$ where all methods work. I will note that large values of the Greed can be needed for MSEC and DIIS to work for these, which the predictive approach automatically generates.
c. To compare with some recent all-electron numbers, slightly more complicated problems such as the Pd surface considered by Kim et al[102] or the Fe & Cr semi-surfaces considered by Winkelmann et al[103]. These converge a little more slowly with MSR1, may not be very stable with MSGB and do not always behave as well with DIIS. In general these are quite sensitive to inappropriate scaling, as well as complications due to the pseudo-charge (see Appendix 1).
d. Harder problems such as those which are ill-conditioned or where the atomic positions and densities are simultaneously optimized for large changes in the positions; normally MSR1 significantly outperforms alternatives as will be shown for a WFe multilayer and a chemisorbed water case as two examples.
e. A set of thirty-six different structures comparing the MSR1 and DIIS algorithms with full predictive control to calculations with a fixed algorithm greed, comparable to approaches in other codes. These range from structures reported to be hard to converge, to seven with more than a hundred atoms. They cover a wide range of different types of problems, including: two molecules; surfaces of metal including two with adsorbates; surfaces of small-gap semiconductors or larger-gap oxides; a number of problems with many different types of chemical environments; some more routine structures.

In a few cases with MSGB the predictive algorithm crashed due to ghost-bands[104-106]; at the time of writing the default action in Wien2k is to stop when these are detected instead of backtracking. (A private version of the author's does not stop when ghost-bands occur, and handled these cases without problems.) Beyond that the algorithm never diverged. Some of the fixed Greed calculations in §5.6 diverged or crashed with ghost-bands; they are not protected.

Metrics that will be shown in the following graphs are:

1. The L2 Residue per atom, which corresponds to the root mean squared change of the total volume integrated density (L2) in units of electrons/au$^{1.5}$ (density times volume$^{1/2}$), see §2.2 for an analysis of the units.
2. The Greed, $\sigma_n$ as described earlier in equation (11), a dimensionless number.
3. Damping, the parameter $\beta_n$ in equation (11), dimensionless
4. The step trust region as described in §3, in the same units as the residue
5. How much reduction there was as well as the predicted reduction, which is given by

$$Reduction = \|[I - Y_n Inv(Y_n^T W)W^T]Res_n\|/\|Res_n\| \qquad (26)$$

6. How large the step taken was compared to the residue, dimensionless
7. The effective rank, which was defined in equation (19), dimensionless
8. The root mean squared pseudo-force when relevant, in Ryd/au
9. For the last case, the total energy relative to the initial energy in Ryd.

Except for the case of bulk MgO in §5.1 in all cases default parameters are used, where the Greed can range from $10^{-3}$ to 1.0, $1 < \beta_n < 0.15$, there is no upper bound on the maximum step trust radius or atomic movements, although there were lower bounds on these of $10^{-3}$ and $10^{-4}$ au. For the examples in §5.2 and §5.6 the number of prior memories was the default 10, in all other cases where just the density was being converged the number was the default 8.

## 5.1 Bulk MgO

A simple example is bulk MgO without spin polarization for which MSR1 results are shown in Figure 2a. Technical parameters are the PBE functional[107], muffin tin radii of 1.8 au, a 6x6x6 k-mesh with Fermi-Dirac occupancies at room temperature and a plane wave expansion via RKMAX of 7.0. The convergence of the RMS residue is linear, and after the first couple of iterations the trust radius increases to more than 10 and plays no part. This indicates that the algorithm believes that this is a very linear and well-conditioned problems, which it is. The Greed rapidly increases to 0.7-1.0, the latter is the default maximum value. The step magnitude in each iteration also rapidly rise to 0.8-1.1 of the total residue.

Similar results are obtained with MSEC, MSGB whereas DIIS is slower with the same regularization. This is due to the bias towards the initial large residual, which led to an effective rank of ~0.33 compared to ~0.9 for MSR1. Changing the regularization altered this, increasing the effective rank and improving the convergence as shown in Figure 2b; the smaller the regularization, the larger is the effective rank. This illustrates the effect of the rescaling in MSR1, MSEC and MSGB which helps in this case, although the relationship between the condition number of $Inv(Y_n^T W)$ and the convergence rate does not appear to be simple.

## 5.2 Bulk MgOH$_2$, both density and atomic positions

A second example is simultaneous convergence of densities and atomic positions for bulk Mg(OH)$_2$ without spin polarization, shown in Figure 3. Technical parameters are the PBE functional[107], muffin tin radii of 1.6, 1.2 and 0.5 au for Mg, O and H respectively, a 5x5x3 k-mesh with Fermi-Dirac occupancies at room temperature and a plane wave expansion via RKMAX of 2.5.

As shown in Figure 3a, similar to bulk MgO the Greed rapidly rises to values of 0.2-0.4, and the bound on the total step size in electrons/atom rises from 0.03 to ~11, playing relatively little role. Through most of the iterations the atoms are moving by about 0.01 au per iteration, and the total residue is slowly converging; note that since this is an energy minimization the L2 residue is not a precise metric of the progress. In many cases, particularly towards the end, the change in the combined density/position residue is significantly larger than the L2 residue due to soft modes coupling the plane waves and the atomic positions.

A different view of the convergence is shown in Figure 3b which shows how much the L2 reduces in any given iteration as well as what is predicted from the linear model. Sometimes the step is not very good; on four occasions the algorithm backtracks and in one case the step is recalculated.

### 5.3 Transition Metal Surfaces

In order to compare with some recent work, cases shown in Figure 4 are a non-magnetic Pd surface with 15 atoms similar to what was used by Kim et al[102], a non-magnetic Cr and a magnetic Fe surface, both with 19 atoms as recently used by Winkelmann et al[103] and a distorted, icosahedral 45 atom Ru cluster with a nearby N atom used by Woods et al[41, 42]. In all cases the PBE functional[107] was used, and other technical parameters are:

- Pd Surface: RMT 2.5, RKMAX 7.0 with a 10x10x1 k-mesh, Fermi-Dirac occupancy at room temperature.
- Cr Surface: RMT 2.25, RKMAX 8.0 with a 10x10x1 k-mesh, Fermi-Dirac occupancy at room temperature, a cell with a volume of 1620 au$^3$
- Fe Surface: RMT 2.10, RKMAX 7.0 with a 10x10x1 k-mesh, Fermi-Dirac occupancy at room temperature, a cell with a volume of 1589 au$^3$
- Ru cluster: RMT 2.26 and 2.42 for the Ru and Ni respectively with just the gamma-point and Fermi-Dirac occupancy at room temperature.

Comparing results with different codes is not simple. The results for the Cr and Fe surfaces are better than those reported by Winkelmann et al[103] for their calculations without a Kerker preconditioner, an important caveat being that while the units used[52] appears to be very similar, they used a different final metric, dividing by the square root of the volume to give au$^{-3}$, and a convergence tolerance of $10^{-6}$. In terms of the results in Figure 4a this is approximately equal to $10^{-4}$, with caveats when comparing codes. No attempt has been made to optimize the mixing parameters in Figure 5, whereas Winkelman et al[103] report only their best results.

The results for the Pd surface are comparable to those reported by Kim et al[102] when they used a Kerker preconditioner, with the same caveats about differences of units and codes. Kim et al were unable to get their version of MSEC to converge. However, as they admit, their "MSEC" does not contain key components of the original algorithm from 2008[23]. In this authors opinion it is not surprising that an unscaled and unprotected algorithm does not converge, as Kim et al[102] found.

The icosahedral cluster converges herein without any problem, it is a well-conditioned problem. In the work of Woods et al[41, 42] it stagnated near the solution after 500 iterations[67]. This was with a very different plane-wave pseudopotential code (Castep), so the reason why it did not converge merits investigation. Part of this may be because of how the authors controlled the Greed, which they do not describe in their paper beyond stating that they do.

These are cases where the problem is somewhat non-linear, so the Damping term plays an important role. Values of this for just the iron surface are shown in Figure 4b, they are similar in the other cases. For these cases the Damping averaged over the iterations was 0.66, 0.75, 0.65 and 0.91 for Pd, Cr, Fe and the cluster, consistent with how "noisy" the convergence is. In tests, without the Damping the convergence is significantly worse in most cases, as expected.

### 5.4 Tungsten-Iron multilayer

A more complex case shown in Figure 5 is a multilayer containing 14 iron atoms and four tungsten in a 2.87x4.06x35.00 Angstrom unit cell with Cmma symmetry, spin-polarized. Technical parameters are the PBE functional[107], muffin tin radii of 2.0 and 2.31 au for Fe and W respectively, a 7x7x1 k-mesh with Fermi-Dirac occupancies at room temperature and a plane wave expansion via RKMAX of 7.0. The iron atoms started with a magnetic moment of 3 (ferromagnetic), the tungsten 2. (The converged values had the tungsten non-magnetic, with a total spin of 31 within the unit cell.)

This is a less stable problem, and was hard to converge and often crashed with earlier versions of the mixer in Wien2k. The convergence of the L2, shown in Figure 5 follows what one expects:

a. MSR1 converges quite cleanly.
b. MSGB does converge, but is clearly noisy.
c. DIIS is adequate as it focuses on the largest eigenvalues and will converge those.
d. MSEC does not emphasize the largest eigenvalues, and would eventually converge but is very slow in this case.

### 5.5 MgO surface with hydroxide

Another example is a non-spin-polarized calculation for a MgO surface with chemisorbed water and simultaneous optimization of atomic positions, shown in Figure 6. Technical parameters are the PBE functional[107], muffin tin radii of 1.63, 1.20 and 0.60 au for Mg, O and H respectively, a 3x1x1 k-mesh and a plane wave expansion via RKMAX of 2.5. Due to rotation of the hydroxide this has significant soft modes.

The energy convergence, shown in Figure 6a is as expected; both MSR1 and MSGB do well with MSGB being somewhat noisy; MSEC and DIIS come close but are not so good with the total energy -- they are less effective with the soft modes. This is clearer in Figure 6b which shows the total movement in au of the hydrogen atom (number 3) that moves most for the four different algorithms; both DIIS and MSEC move it less. The difference is clearer in Figure 7 which shows the top of the initial structure on the left, the MSR1 result and the DIIS result. A vertical, dashed line is drawn through atom H3 to guide the eye. Both MSR1 and MSGB are able to handle the soft mode associated with the shears indicated, allowing H3 to be more strongly hydrogen bonded.

### 5.6 Comparison to fixed greed cases

To analyze the effect of the predictive controls, I will compare the results with the full algorithm to those where a fixed algorithm greed is used, the Damping and Trust region of §2.4, the backtracking of §2.5 and the predictions of §3 have all been removed. The units of §2.2 as well as the regularization of §2.3 and the scaling of §2.6 have been retained as removing these is inappropriate. These fixed greed calculations are similar to what is in most other codes, where the user might have to change the greed if they do not converge, or perform other adjustments.

A total of thirty-six different structures are included herein, ranging from somewhat simple to relatively complex. These include insulators, metals, surfaces, a couple of on-site hybrids, a couple of molecules as well as a number of large more complex structures. Very hard problems such as the WFe described earlier are not included, as they diverged for the range of fixed values used herein. (Many other cases also diverge with fixed Greed unless it is very small.) The intent here is to test more complex cases including ones with a significant number of different types of atoms and chemical environments. The structures can be put into four groups:

1. Structure such as $Cr_2$ and $CrC$ which were discussed by Daniels and Scuseria[108] as being hard, and also a number of structures that Woods et al[42] consider to be hard, which can be found in their depository[109].
2. A number of relatively large and complex structure, including an iron vanadate $SrFe_3V_{18}O_{38}$,[110] a large unit cell titanyl sulfate[111]; a complex silicate[112] $Ba_2Fe\ Ce_2Ti_2Si_8O_{26}$; a large unit cell intermetallic[113] $Au_{10}Mo_4Zn_{89}$ plus two artificial substitutional derivatives of it $Au_{10}Ga_4Mo_4Os_{12}Ru_4V_{10}Zn_{49}$ and $Ag_6AlAu_4Cr_6Cu_{12}Fe_{12}Ga_4Mo_4Nb_{12}Ni_4Os_{12}Ru_4V_{10}Zn_{12}$; a carbonyl $Pb(Mn(CO)_5)_3(AlCl_4)$[114]; a small band-gap semiconducting silver-ion conductor $Ag_8TiS_6$[115], a magnelli phase superstructure[116] and a quasicrystal approximant $Al_{34}Ni_{11}$[117].
3. A number of different surface structures, including two experimentally-relevant structures for $SrTiO_3$ (110) surfaces[118] and one for BN on Rh[119]; two semiconductor small band-gap (001) surfaces for SnSe and SnTe; an Al (001) surface; a $BaCuF_4$ (001) as well as carbon on a Ni (001) surface from Woods et al[42].
4. More conventional cases included a couple of structures from the examples that comes with the Wien2k code, and a few others collected from various other sources over the last decade.

In many cases fixed Greed and DIIS failed to converge simultaneous atom and density problems, so all of the calculations are for fixed positions.

The average number of iterations to converge for these test cases with the predictive algorithm was 26 for MSR1, 29 for MSEC, 31 for MSGB and 32 for DIIS. None of them were problematic for the predictive approach, although a fair number of the fixed-value calculations either diverged or did not converge within a reasonable number of iterations (typically 200).

The results relative to the full algorithm MSR1 are shown in Figure 8 for fixed-value MSR1, and Figure 9 for fixed-value DIIS, with addition information in Supplemental Table S1; all structures are included as CIF files in the Supplemental Material, and technical parameters for the calculations are embedded in the CIF files. In both cases the raw ratios for different Greeds are plotted on the left, and a histogram of values on the right. Cases where the ratio was larger than 3 have been added to that in the histograms, as well as those that did not converge with the fixed Greed. The change if the fixed-greed results for the Pd, Fe, Cr and WFe cases were included is shown in red, as these only converge for very small fixed Greed.

Figure 10 shows the average ration as a function of Greed for MSR1 and DIIS, where the cases that did not converge have been included with three times the number of iterations. The predictive algorithm on average does better than the fixed ones, particularly as it did not crash or diverge.

The calculations without the controls fall into several classes:

1. Cases where larger Greed converges faster. As mentioned earlier, this depends upon the magnitude of the non-linear component.
2. Cases where and intermediate Greed is better, and they may diverge for larger values
3. Cases where only a very small Greed works, for instance the Pd, Fe, Cr and WFe cases shown earlier (not included in Table S1 or the graphs in Figures 8 and 9). This is more often the case for metallic or small-gap surfaces.

The number of iterations to convergence did not scale in a simple fashion with the number of unique or total atoms; as described previously[27], it is known[12, 13, 120] that the convergence of quasi-Newton methods depends upon the number and width of eigenvalue clusters. For instance, simple SmS with 2 atoms in the unit cell and active 4f electrons converged worse than a 2×1 (110) $SrTiO_3$ surface which had 74, or the intermetallic $Au_{10}Mo_4Zn_{89}$ which has 103. In certain cases the convergence did not smoothly vary with the fixed Greed, which is an indicator of ill-conditioning, for instance the silicate. There was no major indication that metal systems (i.e. those with partial occupancy at the Fermi energy) converged worse than insulators, and no need to use large smearing terms.

In some cases specific fixed Greed values were a little faster than the predictive algorithm. The prediction is not perfect as it is extrapolating past performance to the future. I will argue that the fact that the algorithm achieves close to optimal speed across a significant range of material system and problems of different stability and non-linearity without any user intervention makes it useful. The histograms in both Figure 8 and 9 indicate that the predictive algorithm is doing well.

## 6. Discussion

The aim herein is a fast algorithm that will always converge, which works in all cases without adjustment. The method has to be applicable to not just density, but also combined density and atoms, orbital potentials, meta-GGAs and anything else. (The code in Wien2k can also handle linearization energies, and an experimental option will "mix" constraints on atomic positions.) While nothing can save a very badly posed problem, the mixing algorithm has to be able to handle both simple and complicated problems, ones which are well-conditioned as well as those which are ill-conditioned. The algorithm should be based upon a combination of solid mathematics and physics/chemistry, as otherwise it is unlikely that it will be generally applicable.

The algorithm herein comes very close to achieving this. If a bad initial density is chosen it is possible that the iterations can crash at the start with ghost-bands[104-106], sometimes in the first iteration. Without prior history it is impossible to safeguard the algorithm at the start, hence the first Pratt-like step is normally small, being cautious.

The different variants behave as one would expect from the mathematics. In general MSR1 is the best, then MSEC, MSGB and DIIS (all with the predictive method). MSGB is greedier, which also means that it is noisier, so it is more likely to fail although it did not for any of the test cases herein.

When atomic positions are included MSR1 is significantly better, with MSGB second, both MSEC and DIIS can fail to properly converge the atomic positions since they do not handle soft modes well.

It should be noted that only the fixed Greed calculations of §5.6 are somewhat "standard" calculations, although since they include proper variable units including multiplicities (§2.2) and regularization (§2.3) they are not the same as those elsewhere in the DFT literature. My experience is that incorrect units and regularization do not matter for well-conditioned problems, but do for ill-conditioned ones.

It is worth reiterating some of the buried points and potential scientific myths of mixing that have been mentioned earlier:

1. No method is guaranteed to converge unless it is safeguarded by either a line search or a trust region control, either implicit or explicit.
2. The different core methods, i.e. Anderson, Pulay, Broyden or other are fundamentally the same except for hidden scalings and regularization that are often incompletely documented. In many cases they all converge the same within stochastic variations so long as the controlling code is appropriate – that is the case herein. I suspect that large differences in the convergence as a function of which algorithm is used reported in the literature are in many cases due to buried parameters.
3. Frequently poor convergence is due to a poorly posed physical problem. Nothing says that every DFT problem should have a simple, unique fixed-point solution, and many certainly have multiple local fixed-points. Empirically the radius of convergence of DFT problems depends upon how well posed they are in terms of the underlying physics and chemistry. While this is hard to prove rigorously, it is common in practice.
4. Phenomena such as charge sloshing are physical manifestations of non-linearities of the mathematics, ill-conditioning or perhaps inappropriate algorithms or problems.
5. Large Greed (mixing factors) are not necessarily good or desirable.
6. Small Greed can be as much of a problem as large Greed.
7. With limited memory methods for optimization only a few memories are typically used, more is slower – the same herein. The effect of numerical and algorithmic noise will scale inversely with the square root of the number of memories used. Hence, if a large number is needed this suggest possible problems.
8. While simple implementations of mixing methods will probably work, one can do much better with attention to the scaling, units and the underlying mathematics.

The algorithm described herein is more complicated than others in the literature. However, it is no more complicated than sophisticated optimization or non-linear least squares algorithms (e.g.[11, 13]). Generating code to solve the trust region subproblem and the predicted parameters is not that complicated, and once done extensions to other trust controls is trivial. Additional trust controls are available in Wien2k, for instance on the maximum change of any plane wave component; however, tests indicate that they are redundant and degrade performance. To go between the different algorithms only requires changes of the $W$ matrix. To what extent MSR1 and also simultaneous density and atom movements will work with pseudopotential codes is not yet clear since it has never, to my knowledge, been tested with modern fixed-point methods.

Values for some parameters such as the regularization may be code dependent. Various parts of DFT codes involve numerical differentiations, integrations and quadratures. These, and other parts of the underlying DFT code can lead to ill-conditioning and instabilities; details are rarely if ever reported. This importance of algorithm "noise" and instability was previously mentioned by the author[27], is well known in optimization[15], and has been analyzed in a different context by Toth et al[121]. Over the last few years some attention has been paid to improving these in the Wien2k code, which has contributed to improved speed in terms of the number of iterations to convergence.

As a specific example, numerical differentiation is often performed using a polynomial fit over values. As the degree of the polynomial increases the fit becomes more accurate. However, the conditioning becomes worse. To avoid this all one-dimensional numerical differentiations in the Wien2k have been converted to use a cubic spline fit, which is much better conditioned. A minor loss of accuracy (at the $10^{-8}$ level or less) is more than compensated for by less iterations to convergence and more general stability, particularly since there are always other hidden approximations in DFT codes. Particularly when densities and atomic positions are simultaneously converged, noise due to ill-conditioned internal algorithms or numerical truncations can play a significant role in degrading performance.

It is not hard to combine the methods described herein the various forms of Kerker estimation of the Jacobian or its inverse (e.g.[50, 90, 102, 103, 122-128]). There are clear indications that this improves convergence, although there is more to this than appears to have been discussed to date. Two different approaches that can be used. The first is to estimate the initial matrices, rather than using a unitary matrix, i.e. change equations (11) and (13) to

$$H_n = \sigma_n H_n^{Est}[I - Y_n Inv(Y_n^T W) W^T] + \beta_n S_n Inv(Y_n^T W) W^T \qquad (27)$$

$$(\rho_{n+1}, R_{n+1}) = (\rho_n, R_n) + H_n Res_n = \sigma_n H_n^{Est} U_n + \beta_n P_n \qquad (28)$$

where $H_n^{Est}$ is the estimate. Note that this changes the unpredicted component, but has no effect on the predicted. Better estimating of the unpredicted step should always improve convergence, and will not change the trust region algorithms significantly although it will change the contribution of the unpredicted to the L2 metric. In principle the initial estimate for the density could be combined with estimation of the Hessian from a simple spring model (e.g. [129]) when refining atom positions at the same time.

The other approach redefines the secant equation (10) as

$$H_n Y_n = \Phi_n^{-1} H'_n \Psi_n Y_n = S_n \qquad (29)$$

Where $\Phi_n^{-1}$ and $\Psi_n$ are matrices (or operators) chose such that $H'_n$ is better conditioned, ideally diagonal. This leads to the change of variables to new $Y'_n$ and $S'_n$ given by

$$Y'_n = \Psi_n Y_n \text{ and } S'_n = \Phi_n S_n \qquad (30)$$

With $Y'_n$ and $S'_n$ then used to replace $Y_n$ and $S_n$ in the equations/algorithms. This changes not just the unpredicted component, but also the predicted component, the relative fraction of previous steps in the MSR1 algorithm and the effect of regularization. Depending upon how $\Phi_n^{-1}$ and $\Psi_n$ are chosen the algorithm will change from a version of "bad Broyden" to Greedy "good Broyden" or a hybrid similar to MSR1.

Note that by using a physics-based definition in equation (30) of the secant condition one automatically is converting from possibly somewhat arbitrary units to ones which are correctly scaled in the sense of section 2.2. To what extent Kerker conditioning in all-electron and other codes remedies inappropriate scaling is an open issue that merits attention.

Methods for Kerker estimation have recently been reported for all-electron codes[102, 103] and were compared in Figure 4; at the time of writing no Kerker estimation code has been constructed for the Wien2k code, so it is not plausible to test the different approaches in detail, particularly for stiff problems where trust region control is essential. I am not aware of testing with different forms of the preconditioning of equations (26) and (27), and sometimes details are missing in publications. More detailed analysis is an issue for future work.

There may be other improvements that can be made, in both the algorithm and how the prediction is done. One example is the tensor approach[58-62] that, similar to the approach here, uses prior steps to estimate the higher-order diagonal terms which is comparable to using an initial estimation.

In summary, the predictive approach combines speed and robustness, and is quite general for MSR1, MSGB, MSEC and DIIS methods of calculating the candidate predicted and unpredicted components, although in general MSR1 is better probably because the vector space is larger and it also has inbuilt positive components which are appropriate for DFT problems. It would be premature to claim that MSR1 and the predictive approach is the best for all possible problems, but it appears to come very close to this. Independent of MSR1, the predictive approach appears to be a powerful addition to available methods.

**Acknowledgements**

I would like to thank Lyudmila Dobysheva, Bouafia Hamza, Luis Ogando and Yundi Quan for testing the algorithm herein. I would also like to thank Robert Schnabel for some background information. I am indebted to Peter Blaha for many comments over the years including extensive testing of this and other versions of the mixer in Wien2k. This work was supported by the National Science Foundation (NSF) under grant number DMR-1507101.

| Name | $W$ Matrix | Centering | Form | ReScaling | Notes |
|---|---|---|---|---|---|
| Good Broyden | S | Sequential | Overwriting | None | Rare |
| Bad Broyden | Y | Sequential | Overwriting | None | Rare |
| DIIS | Y | Current | Matrix | None | Common |
| MSEC | Y | Current | Matrix | Diagonal | Obsolete |
| MSGB | S | Current | Matrix | Diagonal | Noisy |
| MSR1 | Y+αS | Current | Matrix | Diagonal | Optimal |
| HYB1 | Y+αS | Current | Matrix | None | Good |
| HYB2 | Y+αS | Sequential | Matrix | Diagonal | Good |

**Table 1:** Different methods of forming the predicted and unpredicted steps, in terms of the projection matrix *W*; whether the iterations include differences between adjacent iterations (sequential) or are about the current point; whether the least-squares problem is solved by overwriting adjacent points or simultaneously for all, and whether there is any scaling. Results are reported here for the DIIS, MSEC, MSGB, MSR1 algorithms. Two slightly different versions HYB1 and HYB2 have also been tested, but since they are slightly worse than MSR1 they are not discussed further herein. For reference, larger values of regularization are needed for the sequential versions.

**Figure Captions**

Figure 1. Illustration of the effect of non-linearities, with the x axis some variable and the y axis the residue. The solid lines 1-3 schematically represent different types of problem. At the origin the gradient is shown in black dashed. Moving to 0.5 along the x axis is good for small non-linearities as in curve 1, not so good for 2 and 3. In addition, the Simplex gradients (red dashed) change drastically.

Figure 2 In a), plot of various parameters and trust radii for bulk MgO using MSR1; in b) a comparison of the effective rank using MSR1, and also DIIS with different regularization as indicated from $2 \times 10^{-4}$ (the default in all cases) down to $10^{-8}$. With MSR1 the convergence is rapid and the trust radius plays no role with a large predicted Greed. With DIIS the default regularization is too strong, and better performance is achieved with smaller values, although this can lead to instabilities.

Figure 3 Plot of various parameters and trust radii for bulk $Mg(OH)_2$ with simultaneous density and atomic position convergence, without spin polarization as further described in the text. The metrics in a) show how the L2 residue and the RMS force both decrease, while the various trust region parameters increase. As shown in Figure 3b, the predicted reduction is in general smaller than the actual, which is not unexpected and herein is not a reason to decrease the trust regions, both implicit and explicit.

Figure 4 In 4a), plot of the L2 as a function of iteration using MSR1 for a Pd surface, Cr and Fe ordered vacancies as well as a distorted icosahedral Ru cluster with a nearby N atom as described in the text. The variation of the Damping for the Fe surface is in 4b and the structures are shown in 4c for Pd and Fe/Cr. The Damping plays an important role in stabilizing the fixed-point problem in many cases.

Figure 5 Plot of the L2 residue for a spin-polarized tungsten-iron multilayer using the different algorithms. The MSR1 algorithm converges well, MSGB is noisy but converges, DIIS is having some trouble and MSEC does not do well here although it is still improving when the calculation

was stopped. The structure is shown on the right, with tungsten gold and iron red, and the unit cell is marked.

Figure 6 In a), plot of the energy in eV versus iteration number of a larger MgO surface with chemisorbed water relative to the initial energy. As expected from the math, both MSGB and MSR1 handle the soft modes well while DIIS and MSEC are less effective. In b) the total distance in au moved by hydrogen atom 3, see also Figure 7.

Figure 7: Plot of the top two layers for the MgO+$H_2$O case, with hydrogen atom 3 (red) that shown in Figure 6. The initial structure is on the left, MSR1 final in the middle and DIIS final on the right. Dark blue are Mg, light blue O and grey or red are hydrogen. During the iterations hydrogen bonds are formed (blue, dashed) and there are shears indicated by the red arrows. The vertical lines are through atom 3, to help visualize the differences.

Figure 8: Results for the ratio of the number of iterations to convergence using MSR1 without prediction and other controls as described in the text, to that with the MSR1 algorithm and prediction as well as controls. On the left the ratio versus Greed is shown, with the region where the ratio is less than one shaded for clarity. The numbers are in Table S1 of the Supplemental Material, and values that diverged have not been included. On the right a probability histogram is shown, where cases that did not converge are added to the 3.0 number. In red is how the results would change if non-convergent cases for the Pd, Fe, Cr and WFe cases were included. Atom positions are in the CIF files of the Supplemental Material, with technical parameters embedded in the files.

Figure 9: Results for the ratio of the number of iterations to convergence using DIIS without prediction and other controls as described in the text, to that with the MSR1 algorithm and prediction as well as controls. On the left the ratio versus Greed is shown, with the region where the ratio is less than one shaded for clarity. The numbers are in Table S1 of the Supplemental Material, and values that diverged have not been included. On the right a probability histogram is shown, where cases that did not converge are added to the 3.0 number. In red is how the results would change if non-convergent cases for the Pd, Fe, Cr and WFe cases were included. Atom positions are in the CIF files of the Supplemental Material, with technical parameters embedded in the files.

Figure 10: Average ratio relative to the predictive approach as a function of the Greed, where non-convergence has been included as a value of 3.0.

**Appendix 1: Pseudo-Charge**

There is a specific issue with all-electron muffin-tin methods that merits description, as it does not appear to have been analyzed in the literature. With these methods the basis sets are spherical harmonics within the muffin tins, and plane waves outside. While the plane waves are formally only involved outside the muffin tins, they are not automatically zero inside them; setting them to zero would introduce ringing at high Fourier coefficients which can lead to convergence complications. When mixing is being performed the convergence of the components inside the muffin tins is included in the matrices, even though it formally plays no role in the Hamiltonian.

(Numerically it is not so straightforward since it plays a role in numerical derivatives near the edge of the muffin tins.) I call this plane wave density the "Pseudo-Charge".

The default starting density is a summation of single atom densities with an extrapolation of the plane waves within the muffin tins. This extrapolation is in general somewhat different from what is present when the iterations are converged. Unless care is taken the pseudo-charge can drive the mixing; densities where the pseudo-charge is converged but the "real" density is not can easily occur and lead to problems. The limitations on the total step control this. An additional numerical approach is taken in the first iteration, projecting the new density within the muffin tins and using the component of this which is close to the origin in the next iteration. This avoids numerical issues with derivative discontinuities or similar near the muffin tin boundary, while compensating for most of the pseudo charge which, with atomic densities, is near to the nucleus. There may be better approaches for generating the initial density to further reduce these problems, a topic for future work.

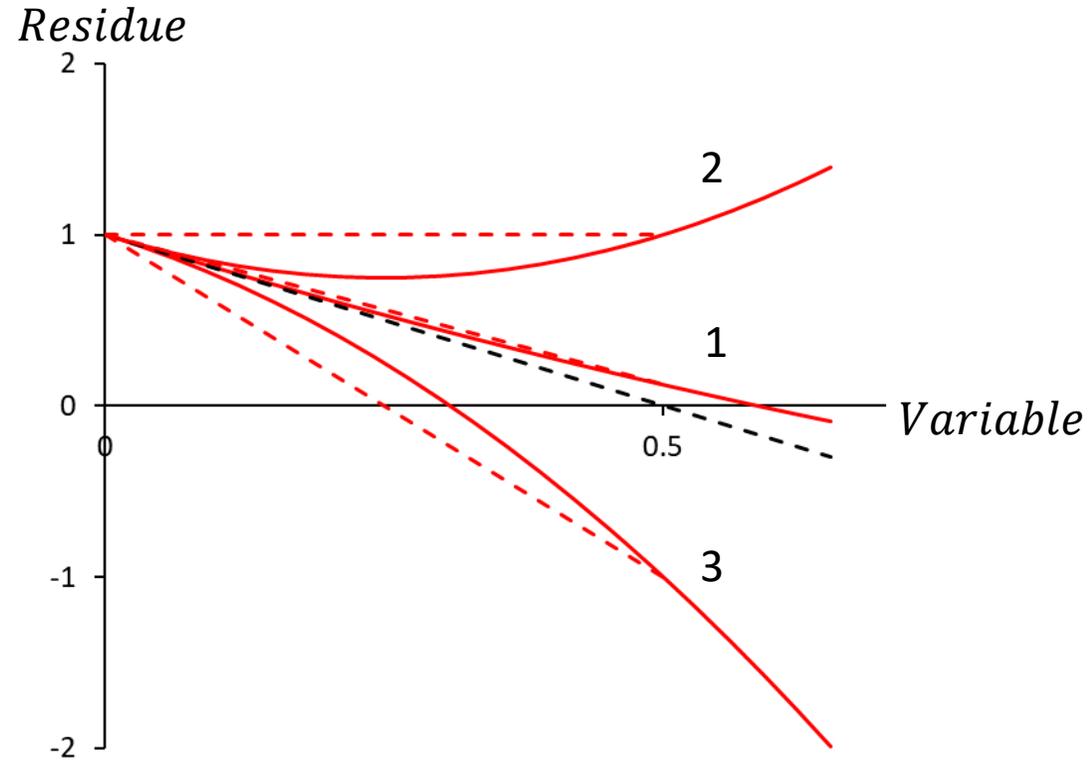

Figure 1. Illustration of the effect of non-linearities, with the x axis some variable and the y axis the residue. The solid lines 1-3 schematically represent different types of problem. At the origin the gradient is shown in black dashed. Moving to 0.5 along the x axis is good for small non-linearities as in curve 1, not so good for 2 and 3. In addition, the Simplex gradients (red dashed) change drastically.

Figure 2a

Figure 3b

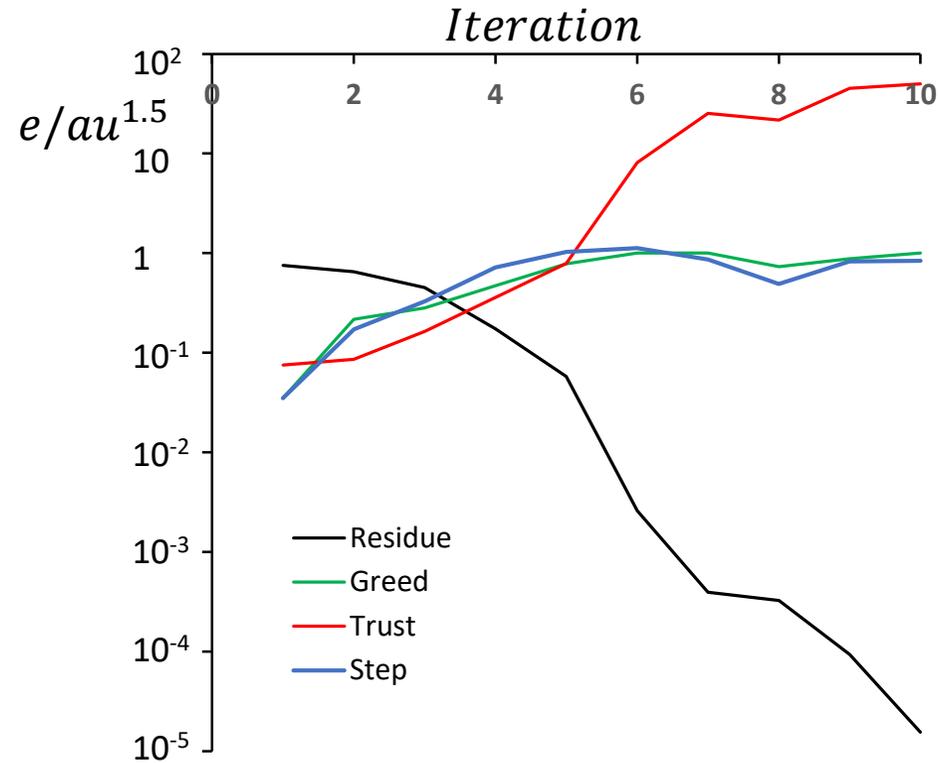
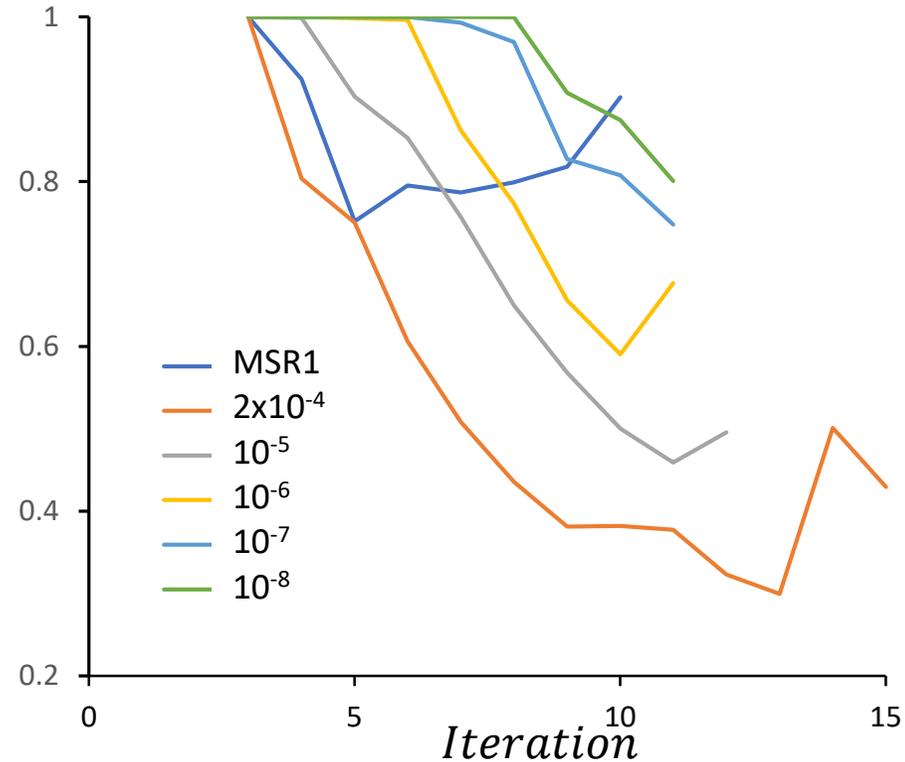

Figure 2 In a), plot of various parameters and trust radii for bulk MgO using MSR1; in b) a comparison of the effective rank using MSR1, and also DIIS with different regularization as indicated from $2\times10^{-4}$ (the default in all cases) down to $10^{-8}$. With MSR1 the convergence is rapid and the trust radius plays no role with a large predicted greed. With DIIS the default regularization is too strong, and better performance is achieved with smaller values, although this can lead to instabilities.

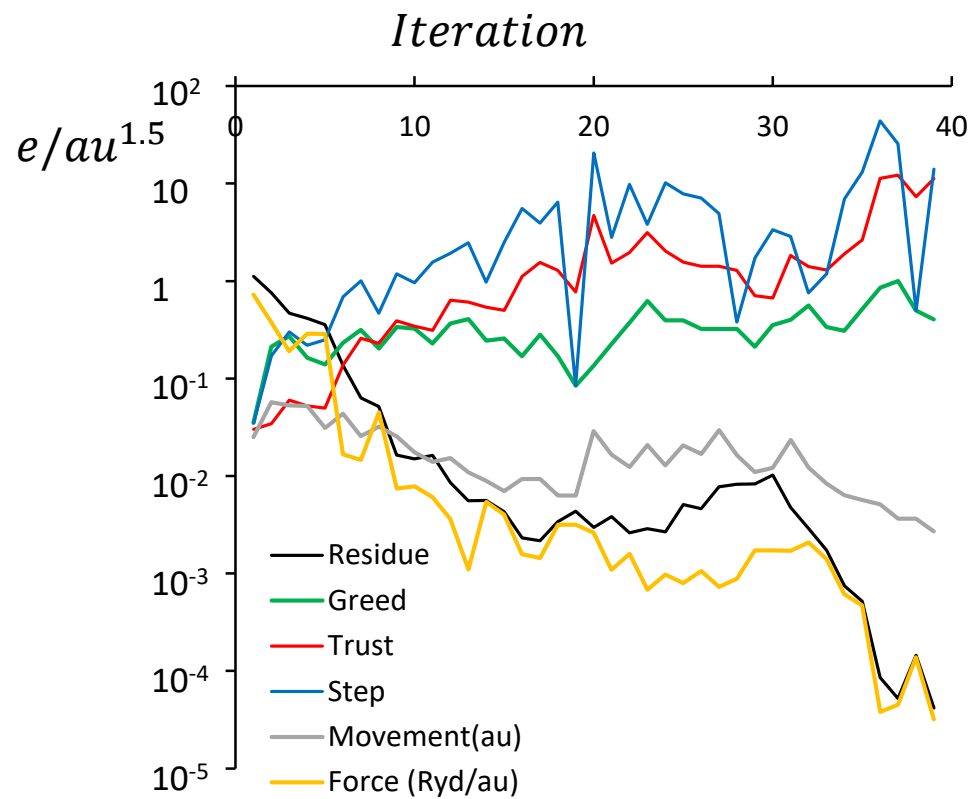
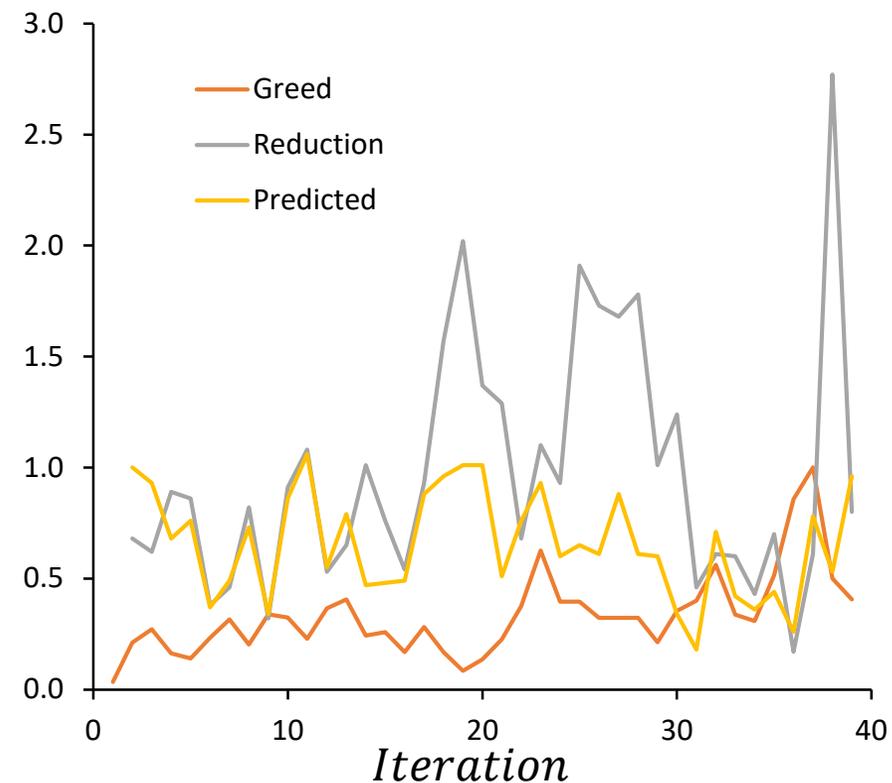

Figure 3a

Figure 3b

Figure 3 Plot of various parameters and trust radii for bulk $Mg(OH)_2$ with simultaneous density and atomic position convergence, without spin polarization as further described in the text. The metrics in a) show how the L2 residue and the RMS force both decrease, while the various trust region parameters increase. As shown in Figure 3b, the predicted reduction is in general smaller than the actual, which is not unexpected and herein is not a reason to decrease the trust regions, both implicit and explicit.

Figure 4a

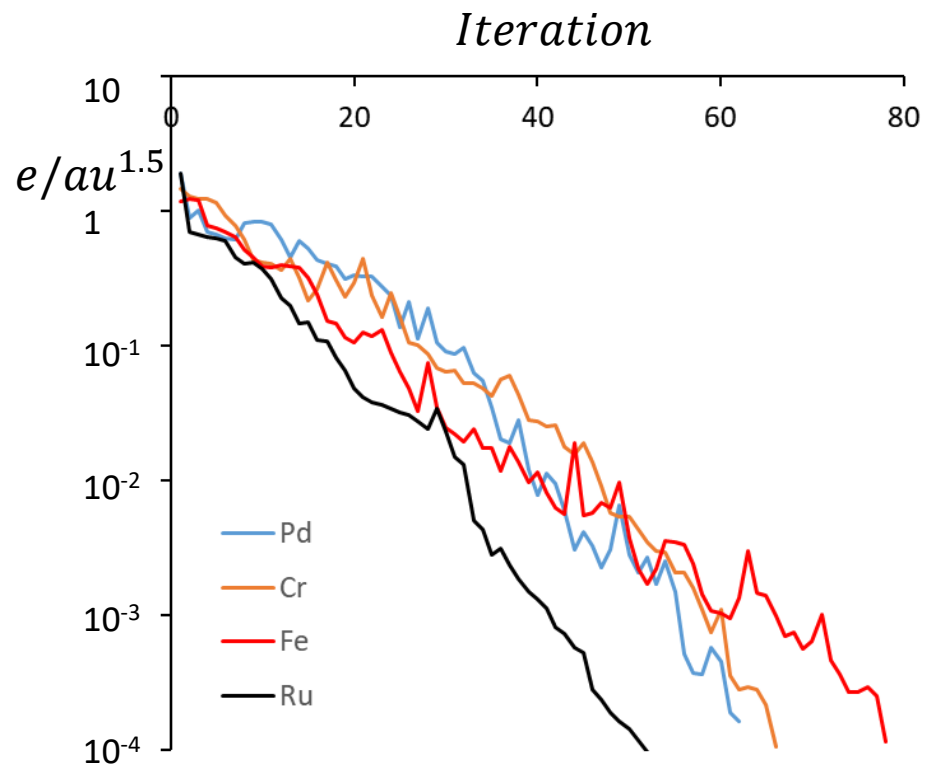

Figure 4b

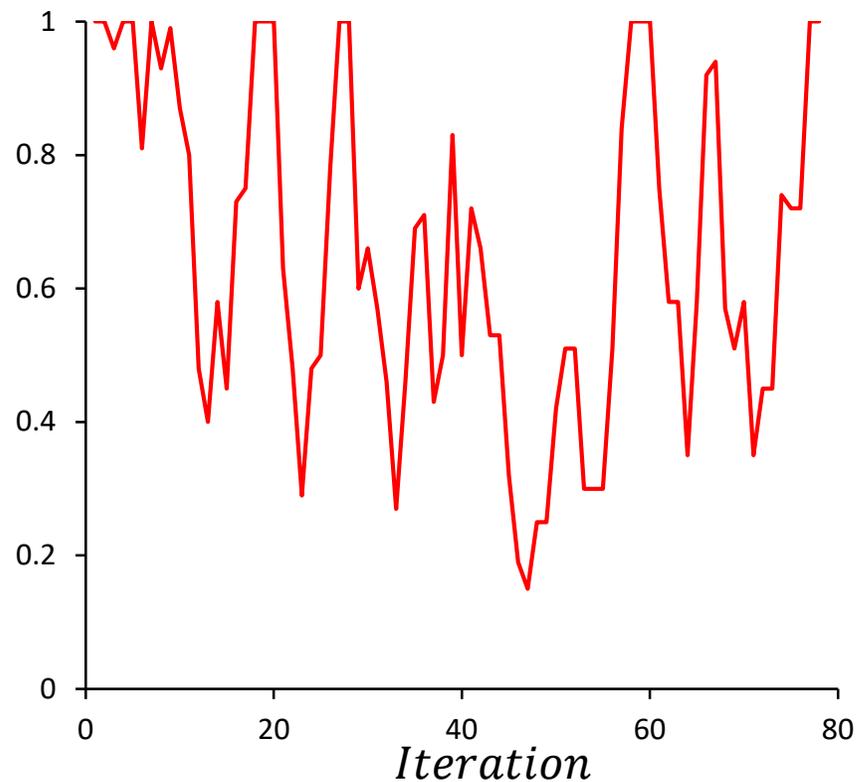

Figure 4c

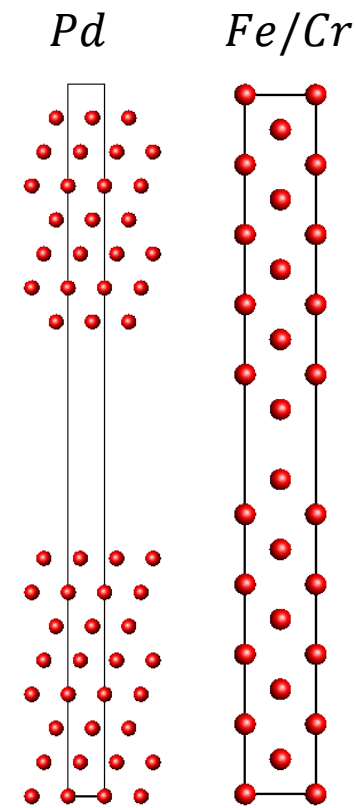

Figure 4 In 4a), plot of the L2 as a function of iteration using MSR1 for a Pd surface, Cr and Fe ordered vacancies as well as a distorted icosahedral Ru cluster with a nearby N atom as described in the text. The variation of the damping for the Fe surface is in 4b, and the structures are shown in 4c for Pd and Fe/Cr. The damping plays an important role in stabilizing the fixed-point problem in many cases.

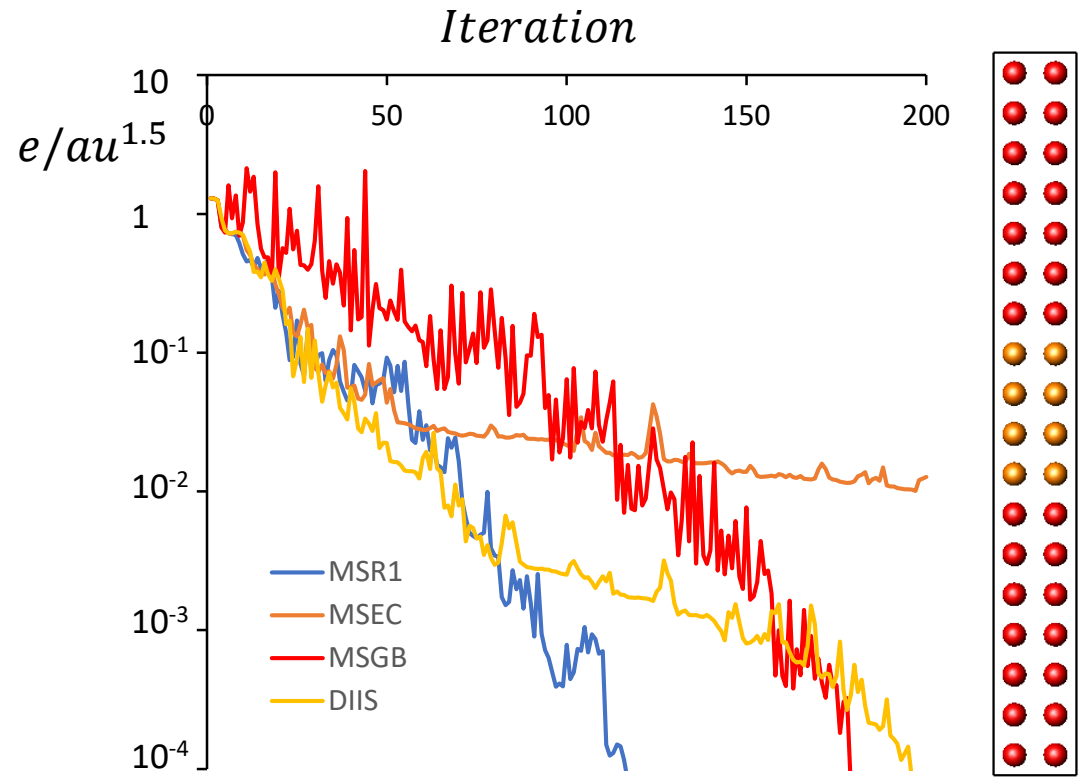

Figure 5 Plot of the L2 residue for a spin-polarized tungsten-iron multilayer using the different algorithms. The MSR1 algorithm converges well, MSGB is noisy but converges, DIIS is having some trouble and MSEC does not do well here although it is still improving when the calculation was stopped. The structure is shown on the right, with tungsten gold and iron red, and the unit cell is marked.

Figure 6a

Figure 6b

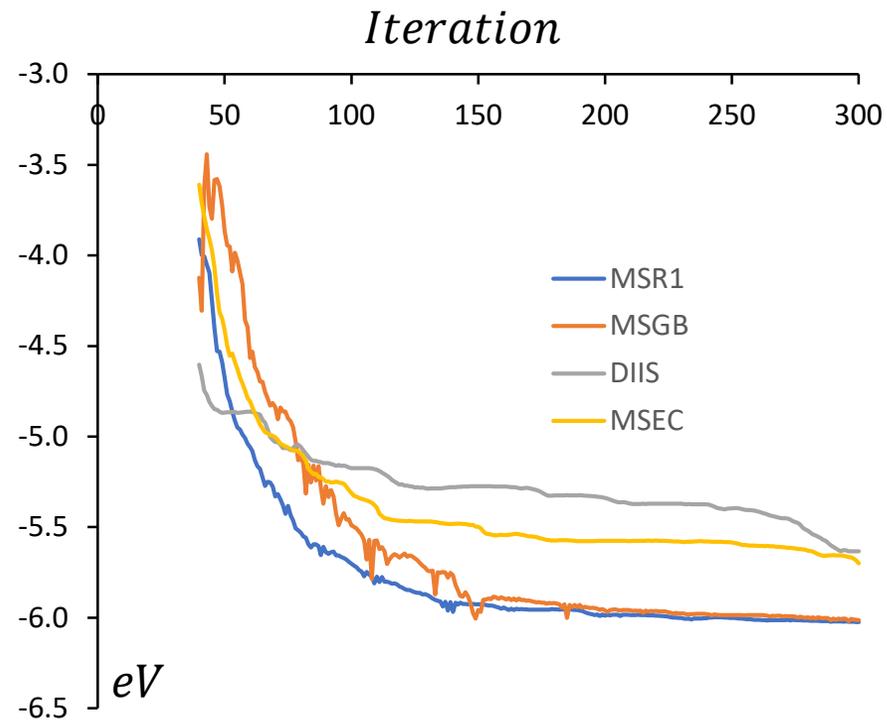
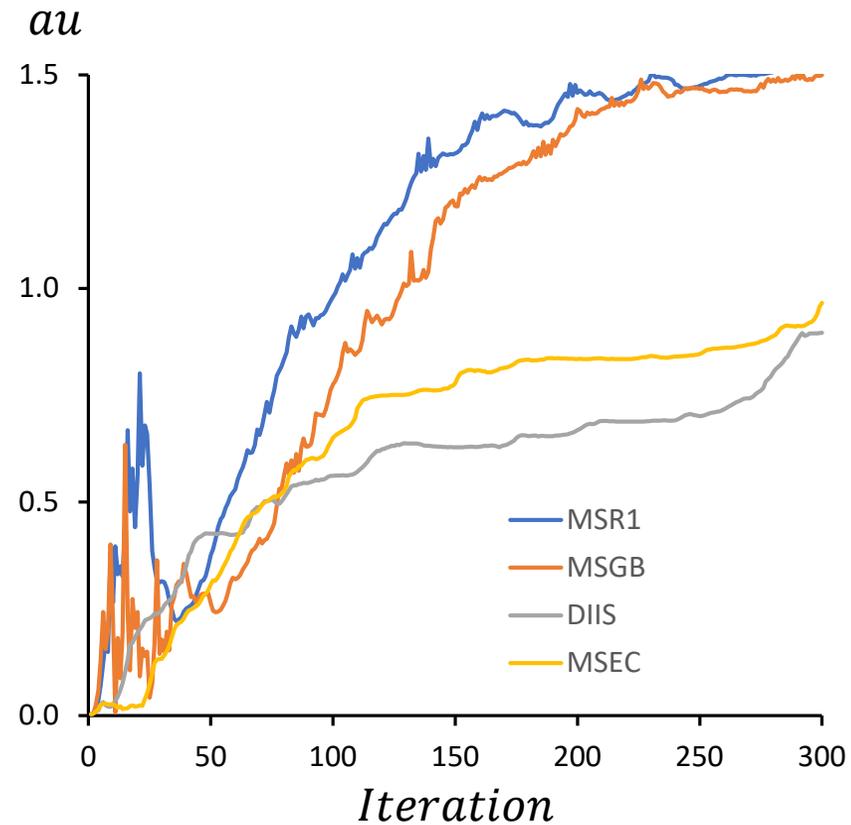

Figure 6 In a), plot of the energy in eV versus iteration number of a larger MgO surface with chemisorbed water relative to the initial energy. As expected from the math, both MSGB and MSR1 handle the soft modes well while DIIS and MSEC are less effective. In b) the total distance in au moved by hydrogen atom 3, see also Figure 7.

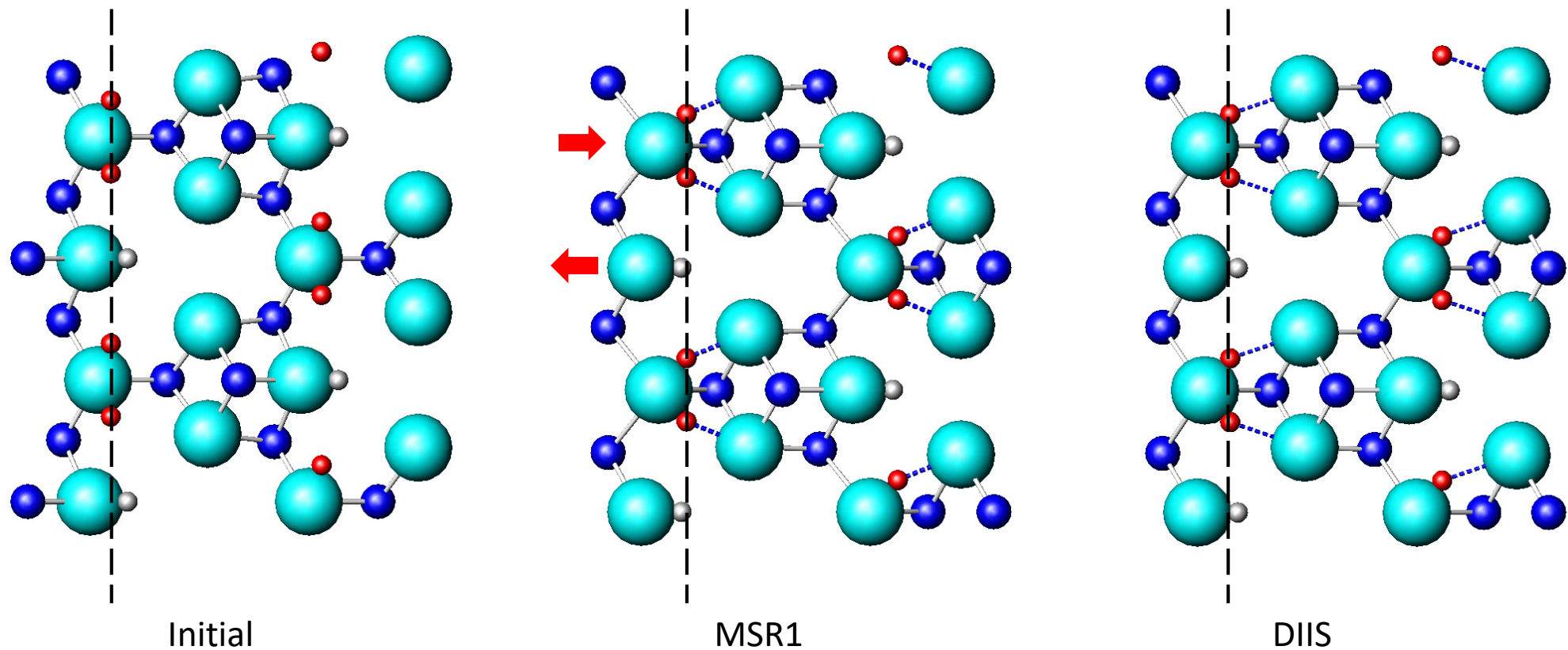

Figure 7: Plot of the top two layers for the MgO+H$_2$O case, with hydrogen atom 3 (red) that shown in Figure 6. The initial structure os on the left, MSR1 final in the middle and DIIS final on the right. Dark blue are Mg, light blue O and grey or red are hydrogen. During the iterations hydrogen bonds are formed (blue, dashed) and there are shears indicated by the red arrows. The vertical lines are through atom 3, to help visualize the differences.

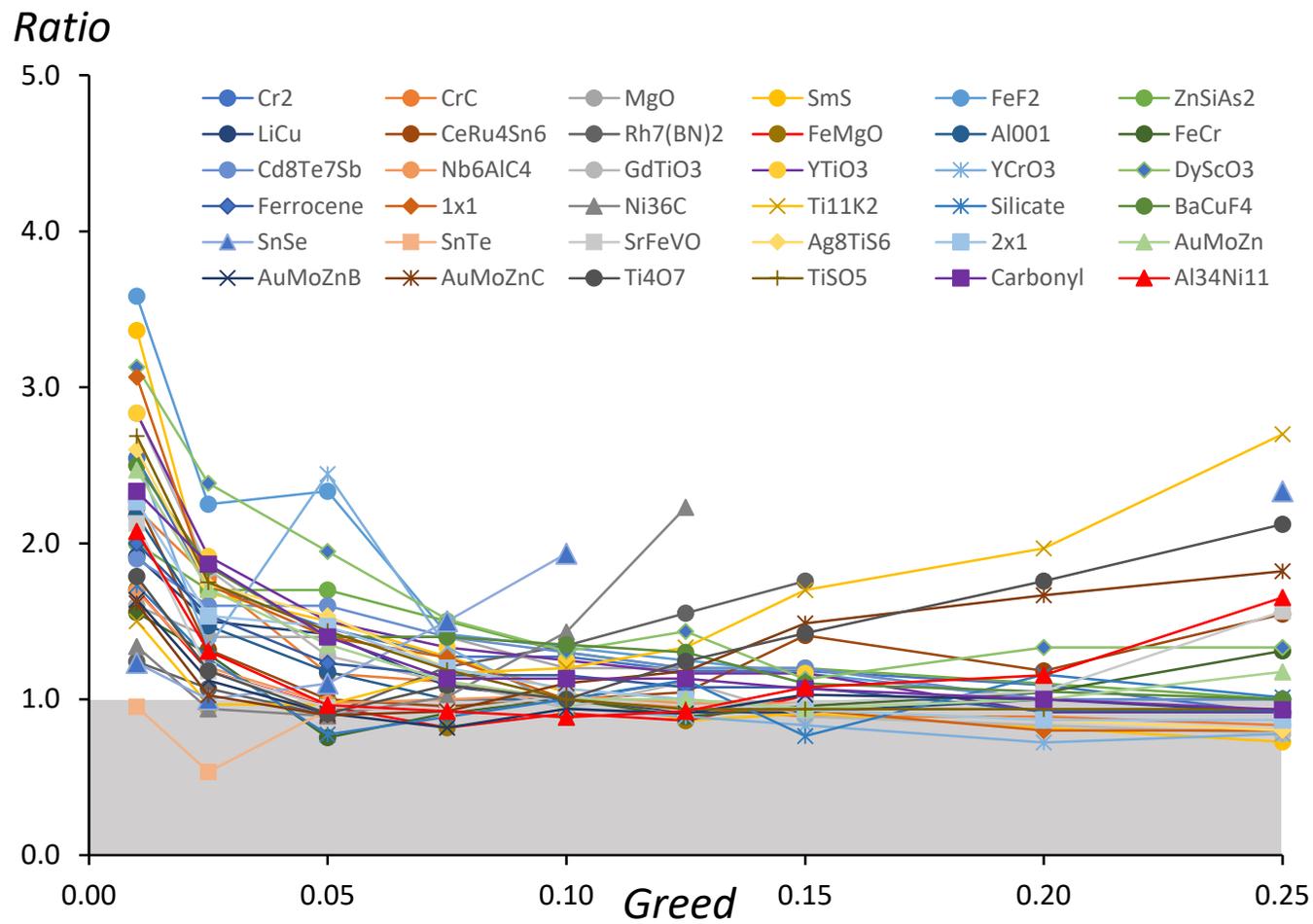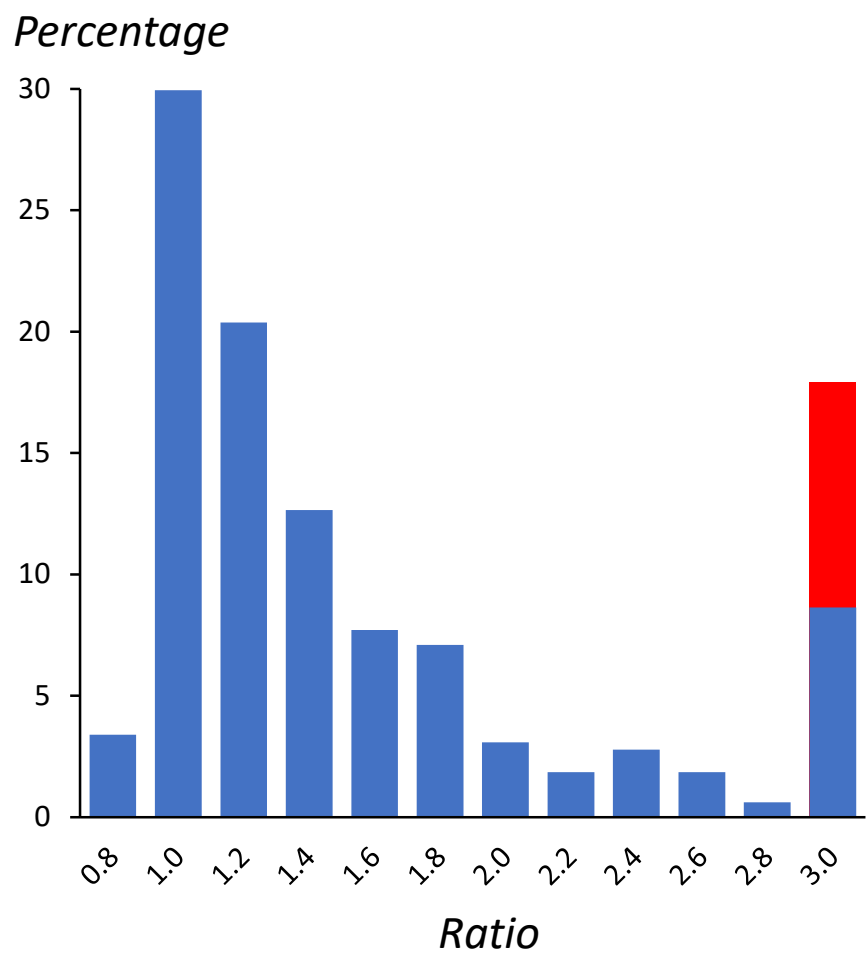

Figure 8: Results for the ratio of the number of iterations to convergence using MSR1 without prediction and other controls as described in the text, to that with the MSR1 algorithm and prediction as well as controls. On the left the ratio versus Greed is shown, with the region where the ratio is less than one shaded for clarity. The numbers are in Table S1 of the Supplemental Material, and values that diverged have not been included. On the right a probability histogram is shown, where cases that did not converge are added to the 3.0 number. In red is how the results would change if non-convergent cases for the Pd, Fe, Cr and WFe cases were included. Atom positions are in the CIF files of the Supplemental Material, with technical parameters embedded in the files.

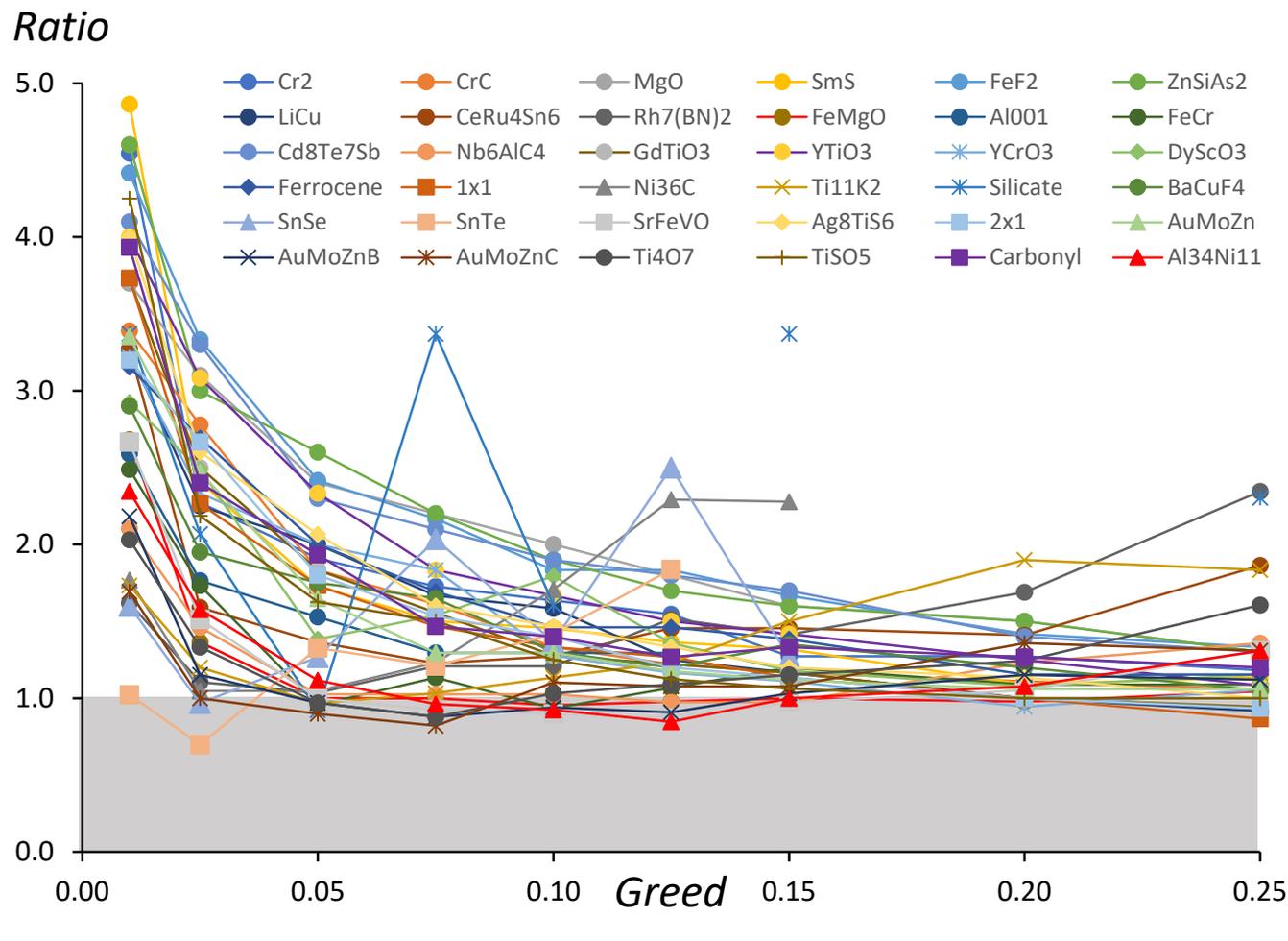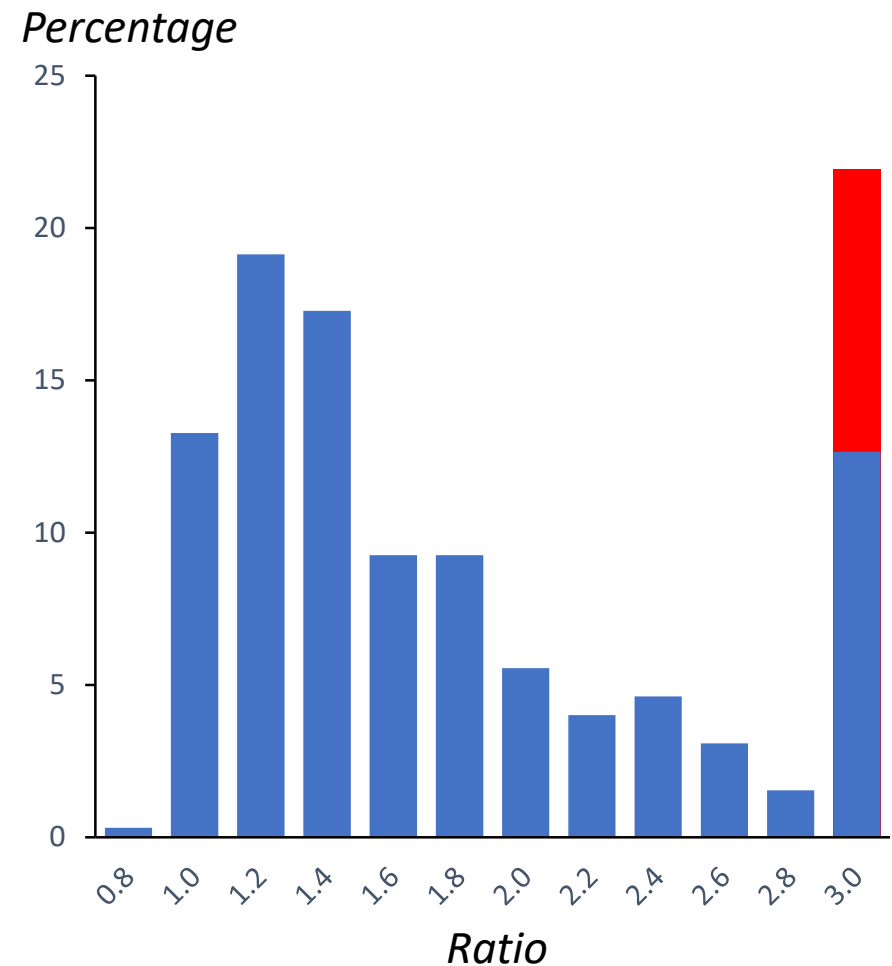

Figure 9: Results for the ratio of the number of iterations to convergence using DIIS without prediction and other controls as described in the text, to that with the MSR1 algorithm and prediction as well as controls. On the left the ratio versus Greed is shown, with the region where the ratio is less than one shaded for clarity. The numbers are in Table S1 of the Supplemental Material, and values that diverged have not been included. On the right a probability histogram is shown, where cases that did not converge are added to the 3.0 number. In red is how the results would change if non-convergent cases for the Pd, Fe, Cr and WFe cases were included. Atom positions are in the CIF files of the Supplemental Material, with technical parameters embedded in the files.

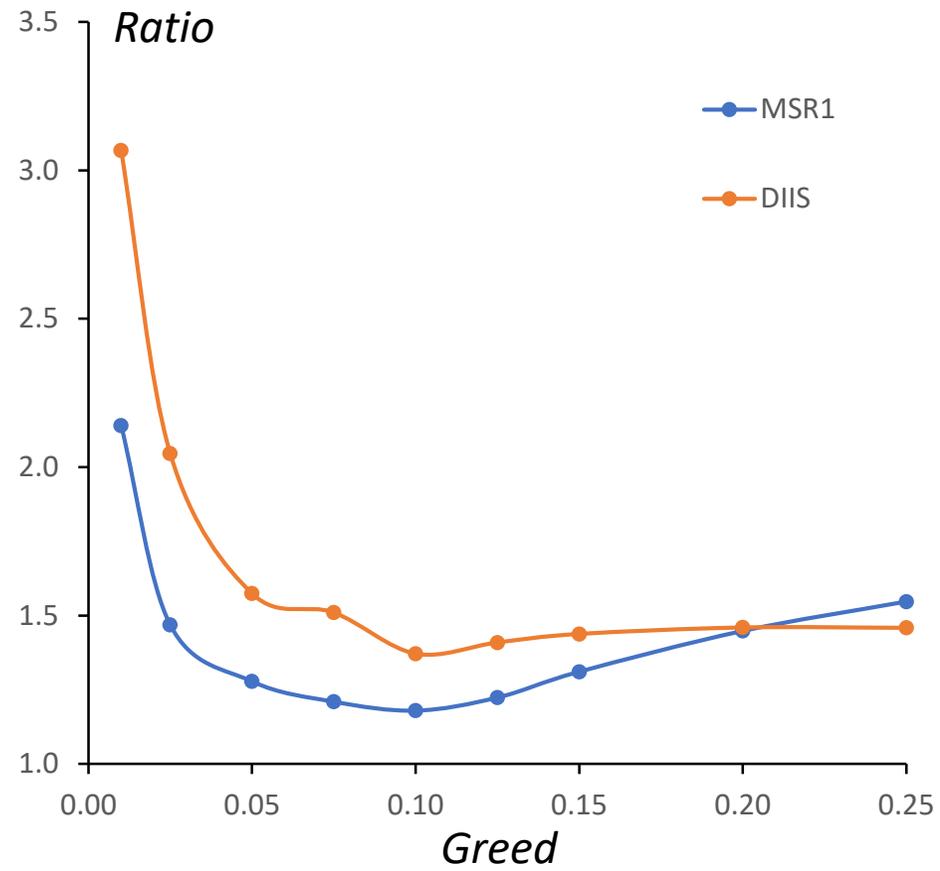

Figure 10: Average ratio relative to the predictive approach as a function of the Greed, where non-convergence has been included as a value of 3.0.